\documentclass[fleqn,usenatbib,useAMS]{mnras}

\usepackage{graphicx}	
\usepackage{amsmath}	
\usepackage{amssymb}	
\usepackage{multicol}        
\usepackage{bm}		
\usepackage{pdflscape}	
\usepackage{scrextend}

\usepackage[T1]{fontenc}
\usepackage{ae,aecompl}

\usepackage{newtxtext,newtxmath}

\title{Global Parameters of Eight W UMa-type Binary Systems}

\author[Poro et al.]{Atila Poro$^{1,2}$,
Mehmet Tanriver$^{3,4}$,
Raul Michel$^{5}$\thanks{Corresponding author: rmm@astro.unam.mx}%
, Ehsan Paki$^{2}$
\\
\\
$^{1}$The Raderon AI Laboratory, Astronomy Department, BC., Burnaby, Canada\\
$^{2}$Binary Systems of South and North (BSN) Project, Tehran, Iran\\
$^{3}$Department of Astronomy and Space Science, Faculty of Science, University of Erciyes, TR-38039, Kayseri, Türkiye\\
$^{4}$Erciyes University, Astronomy and Space Science Observatory Application and Research Center, TR-38039, Kayseri, Türkiye\\
$^{5}$Observatorio Astronómico Nacional, Universidad Nacional Autónoma de México, Apartado Postal 877, Ensenada C.P. 22800, B.C., México\\}

\begin{document}
\label{firstpage}
\pagerange{\pageref{firstpage}--\pageref{lastpage}}
\maketitle

\begin{abstract}
Multiband photometric investigations for eight binary systems of the W Ursae Majoris (W UMa)-type are presented. Six systems are presented for the first time to analyze their light curves. All the analyzed systems have a temperature below 5000 K and an orbital period of less than 0.28 days. We extracted primary and secondary minima from the ground-based observations of these systems. According to a few observations reported in the literature, linear fits were considered in the O-C diagrams, and new ephemerides were presented. Light curve solutions were performed using the PHysics Of Eclipsing BinariEs (PHOEBE) code. The results of the mass ratio and fillout factor indicate that the systems are contact binary stars. Six of them showed the O'Connell effect, and a cold starspot on each companion was required for light curve solutions. Their absolute parameters were estimated and evaluated by two other methods. In this study, the empirical relationship between the orbital period and semi-major axis was updated using a sample consisting of 414 contact binary systems and the Monte Carlo Markov Chain (MCMC) approach. Also, using Machine Learning (ML) and the Artificial Neural Network (ANN) model, the relationship between $P-T_1-M_1$ was updated for a better estimation of the mass of the primary star.
\end{abstract}

\begin{keywords}
eclipsing binary stars, photometry, individual (Eight binary systems)
\end{keywords}


\section{Introduction}
W UMa-type binary systems (EWs) belong to contact binaries, which are considered eclipsing binaries. W UMa-type systems are important astrophysical tools for studying stars' formation, structure, and evolution. Therefore, studying them provides insight into stellar evolution, including key stellar parameters such as mass, temperature, and surface gravity. There are many unresolved issues regarding W UMa systems in our knowledge, despite studies that attempt to answer them (\citealt{1994ASPC...56..228B}, \citealt{2003MNRAS.342.1260Q},
\citealt{2005ApJ...629.1055Y},
\citealt{2007ApJ...662..596L}, \citealt{2008MNRAS.386.1756E},  \citealt{2012JASS...29..145E}, \citealt{2024RAA....24a5002P}). Therefore, it is necessary to model various EW binary stars using observations and to accurately determine the basic physical parameters of the stars. W UMa stars consist of late spectral (A-K spectral type) main sequence stars with orbital periods of less than one day. However, recent studies show that hypercontact binary systems also include M spectral type stars (\citealt{2012AJ....143...99T}, \citealt{2012MNRAS.425..950N}, \citealt{2014ApJ...790..157D}).

In this study, we present photometric observations, light curve solutions, and the determination of astrophysical parameters for eight short orbital period binary systems. These binary systems including ATO J069.8679+53.7711 (hereinafter as J069), CRTS J213545.6+211104 (hereinafter as J213), CSS J004534.6+324435 (hereinafter as J004), CSS J073436.3+290946 (hereinafter as J073), CSS J160934.4+351414 (hereinafter as J160), CSS J214144.0+213748 (hereinafter as J214), NSVS 11234970 (hereinafter as N112) and NSVS 5810460 (hereinafter as N581).

The systems J069, J213, J004, J073, J160, J214, N581, and N112 were included in the Asteroid Terrestrial-impact Last Alert System (ATLAS) catalog (\citealt{2018AJ....156..241H}), and their apparent magnitudes were given in the $g$, $r$, $i$, $z$, and $Y$ bands.
In the \cite{2018AJ....156..241H} study, the classification of each light curve was made by applying the Fourier approximation. So, the variable systems J069 and J213 were classified as contact or near-contact eclipsing binaries\footnote{CBF Close Binary, Full period, ATLAS catalog, \url{https://vizier.cds.unistra.fr/viz-bin/VizieR?-source=J/AJ/156/241}} for which the Fourier fit had found the correct period. Hence, \cite{2018AJ....156..241H} fitted the primary and secondary eclipses separately and variable stars J073, J214, and N112 as contact or near-contact eclipsing binaries\footnote{CBH Close Binary, Half period, ATLAS catalog, \url{https://vizier.cds.unistra.fr/viz-bin/VizieR?-source=J/AJ/156/241}}. The J160 variable is classified as a star that exhibits sinusoidal changes\footnote{SINE sinusoidal changes, ATLAS catalog, \url{https://vizier.cds.unistra.fr/viz-bin/VizieR?-source=J/AJ/156/241}} like ellipsoidal variables. The J004 and N581 variables are classified as stars showing modulated sinusoidal oscillations\footnote{MSINE modulated sinusoidal oscillations, ATLAS catalog, \url{https://vizier.cds.unistra.fr/viz-bin/VizieR?-source=J/AJ/156/241}} that include spotted ellipsoidal variables, rotating stars with evolving spots, and sinusoidal pulsators (\citealt{2018AJ....156..241H}).

These variable systems were also included in the Catalina Surveys periodic variable stars catalog (CSS-CRTS). So, the variable stars J213, J004, J073, J160, and J214 were classified as W UMa-type systems in CSS-CRTS.
By the Adaptive Fourier Decomposition (AFD) method (\citealt{2015MNRAS.446.2251T}) the average $V$ band magnitude was found to be $V=15.84^{mag}$ and orbital period 0.247156-day for J213; $V=14.28^{mag}$ and orbital period 0.246336-day for J004; $V=14.83^{mag}$ and orbital period 0.244346-day for J073; $V=14.70^{mag}$ and orbital period 0.23439-day for J160; $V=14.87^{mag}$ and orbital period 0.240183-day for J214 (\citealt{2014ApJS..213....9D}).

Besides, N581 was classified as a W UMa-type in the Northern Sky Variability Survey (NSVS) variables automated classification catalog (\citealt{2009AJ....138..466H}). Furthermore, in the \cite{2020MNRAS.497.3493Z}, and \cite{2015RAA....15.1493K} studies, N112 was classified as an Eclipsing Binary (EB), and its $V$ band magnitude is given as $13.66^{mag}$.

Since statistics for contact binary systems with periods of about a quarter of a day are quite poor, we chose to observe and study such contact binary systems. Moreover, analyzing more systems may result in a large sample for further investigations in the field of contact binary systems. Applied methods for estimating the absolute parameters of the systems without spectroscopic data were taken into account.

The structure of the paper is as follows: Section 2 provides details on photometric observations and a data reduction process. Section 3 presents extracting minimum times and the updated ephemeris for each system. The light curve solutions for all eight systems are included in Section 4, and the estimation of absolute parameters is presented in Section 5. Finally, Section 6 contains the conclusion.


\section{Observation and data reduction}
The eight contact binaries were observed at the San Pedro Martir (SPM) Observatory in Mexico, which is located at longitude $115^{\circ}27'49"$ West, latitude $31^{\circ}02'39"$ North, and altitude 2830m. These target systems were observed during the course of 18 nights between 2016 and 2022. A 0.84-meter Ritchey-Chretien telescope (f/15), a Mexman filter wheel, and a Spectral Instruments CCD detector (e2v CCD42-40 chip with $13.5\times13.5\mu2$ pixels, gain of $1.39 e-\diagup$ ADU and readout noise of $3.54 e-$) were used for the observations. The field of view was $7.6'\times7.6'$, and binning $2\times2$ was employed during all the observations. The $B$, $V$, $R_c$, and $I_c$ filters were used for photometric observations.

Table \ref{tab1} contains the coordinates, Gaia DR3 distance, Re-normalised Unit Weight Error (RUWE) from Gaia, and orbital period for each system. The systems have short orbital periods, ranging from 0.23 to 0.28 days. Table \ref{tab2} presents the observational characteristics including the year of observations, the exposure time in each filter, $(B-V)_{system}$, and $V_{max}$. We selected the comparison and reference stars for each system based on their closeness, magnitudes, and similarity in color to the target systems.
However, for some systems, there may be few stars in the field of view, making the selection of comparison and reference stars more difficult. Table \ref{tab3} listed comparison and reference stars along with their specifications from Gaia's synthetic photometry (\citealt{2023A&A...674A..33G}).

All the images were processed using the Image Reduction and Analysis Facility (IRAF\footnote{IRAF is distributed by the National Optical Observatories, operated by the Association of Universities for Research in Astronomy, Inc., under a cooperative agreement with the National Science Foundation.}) routines (\citealt{1986SPIE..627..733T}). Images were bias subtracted and flat field corrected before the instrumental magnitudes were computed with the standard aperture photometry method.
We estimated the airmass with Hardie's formula (\citealt{1962aste.book.....H}) and a Python code based on the Astropy package (\citealt{2013A&A...558A..33A}). AstroImageJ software (\citealt{2017AJ....153...77C}) was used for applying airmass to the light curves and also to normalize the flux.

\begin{table*}
\caption{Coordinates of systems from the Simbad database, Gaia DR3 distance, Gaia RUWE, and orbital period of the systems.}
\centering
\begin{center}
\footnotesize
\begin{tabular}{c c c c c c}
 \hline
 \hline
 System & R.A.(J2000) & Dec.(J2000) & Gaia DR3 $d$(pc) & Gaia RUWE & P(day)\\
\hline
J069	& 04 39 28.3248	& +53 46 16.176	& 700(126)
& 17.791 & 0.2711908 (ZTF)\\
J213	& 21 35 45.8424	& +21 11 04.776	& 942(37)
& 0.901	& 0.2471559 (ASAS-SN)\\
J004	& 00 45 34.6824	& +32 44 35.016	& 448(4)
& 1.003	& 0.2463398 (ZTF)\\
J073	& 07 34 36.3504	& +29 09 46.116	& 785(18)
& 1.089	& 0.2443496 (ASAS-SN)\\
J160	& 16 09 34.4400	& +35 14 14.280	& 477(4)
& 1.039	& 0.2343906 (ASAS-SN)\\
J214	& 21 41 43.9728	& +21 37 46.632	& 771(19)
& 1.127	& 0.2401818 (ASAS-SN)\\
N112	& 19 32 07.8816	& +14 58 27.156	& 313(1)
& 0.941	&  0.25074(27) (\citealt{2015RAA....15.1493K})\\
N581	& 20 53 52.8336	& +47 55 17.688	& 275(39)
& 24.851 & 0.2472266 (VSX)\\
\hline
\hline
\end{tabular}
\item 
*ASAS-SN Variable Stars Database, \url{https://asas-sn.osu.edu/variables/}

**The Zwicky Transient Facility (ZTF) catalog of periodic variable stars

***The International Variable Star Index (VSX), \url{https://www.aavso.org/vsx/}
\end{center}
\label{tab1}
\end{table*}

\begin{table*}
\caption{Specifications of observations performed for the systems in the $BVR_cI_c$ filters.}
\centering
\begin{center}
\footnotesize
\begin{tabular}{c c c c c}
 \hline
 \hline
 System & Year of observation & Exposure time(s) & $(B-V)^{(mag.)}$ & $V_{max}^{(mag.)}$\\
\hline
J069 & 2016, 2021 & $B$(40), $V$(20), $R$(10), $I$(10) & 1.15(2) & 13.98(7)\\
J213 & 2017, 2018 & $B$(20), $V$(100), $R$(60), $I$(60)	& 1.00(2) & 15.86(6)\\
J004 & 2021 & $B$(120), $V$(50), $R$(30), $I$(30) & 1.11(3) & 14.68(5)\\
J073 & 2017, 2018, 2020 & $B$(40), $V$(25), $R$(15), $I$(15) & 1.00(2) & 15.09(9)\\
J160 & 2022 & $B$(40), $V$(25), $R$(15), $I$(15) & 1.13(2) & 15.02(8)\\
J214 & 2018 & $B$(300), $V$(240), $R$(180), $I$(180) &  0.94(2) & 14.95(7)\\
N112 & 2018 & $B$(300), $V$(180), $R$(120), $I$(120) & 1.03(3) & 13.70(10)\\
N581 & 2018 & $B$(40), $V$(20), $R$(15), $I$(15) & 0.97(3) & 12.84(8)\\
\hline
\hline
\end{tabular}
\end{center}
\label{tab2}
\end{table*}

\begin{table*}
\caption{List of comparison and reference stars for the studied systems.}
\centering
\begin{center}
\footnotesize
\begin{tabular}{c c c c c c c c c}
 \hline
  \hline
System & Star type &    RA.(J2000)  &  DEC.(J2000)  &   $B^{(mag.)}$   &   $V^{(mag.)}$   &   $R^{(mag.)}$   &   $I^{(mag.)}$   &   $(B-V)^{(mag.)}$\\
\hline
J069     &     Comparison &    4 39 05.7 &   53 45 16.0 & 14.13 & 12.77 & 11.96 & 11.16 &  1.36 \\
               & Reference  &    4 39 53.1 &   53 45 33.3 & 15.59 & 14.45 & 13.77 & 13.08 &  1.13 \\
J213     &     Comparison &   21 35 46.0 &   21 14 29.8 & 14.69 & 13.89 & 13.42 & 12.96 &  0.80 \\
               & Reference  &   21 35 39.1 &   21 13 56.7 & 15.68 & 14.88 & 14.43 & 14.03 &  0.81 \\
J004     &     Comparison &    0 45 18.3 &   32 45 25.0 & 15.01 & 13.84 & 13.12 & 12.52 &  1.17 \\
               & Reference  &    0 45 24.3 &   32 39 30.3 & 13.89 & 12.78 & 12.18 & 11.63 &  1.12 \\
J073     &     Comparison &    7 34 29.3 &   29 08 07.7 & 15.01 & 14.08 & 13.57 & 13.08 &  0.93 \\
               & Reference  &    7 34 40.8 &   29 11 52.0 & 15.02 & 14.24 & 13.80 & 13.40 &  0.79 \\
J160     &     Comparison &   16 09 43.9 &   35 15 56.0 & 14.87 & 13.86 & 13.31 & 12.76 &  1.00 \\
               & Reference  &   16 09 28.9 &   35 13 52.5 & 15.44 & 14.55 & 14.06 & 13.60 &  0.89 \\
J214     &     Comparison &   21 41 37.5 &   21 40 14.6 & 14.40 & 13.66 & 13.25 & 12.87 &  0.73 \\
               & Reference  &   21 41 50.7 &   21 38 37.4 & 15.90 & 15.26 & 14.88 & 14.50 &  0.64 \\
N112     &     Comparison &   19 32 13.1 &   14 56 21.6 & 14.25 & 13.39 & 12.91 & 12.46 &  0.86 \\
               & Reference  &   19 32 16.8 &   14 59 25.0 & 15.76 & 14.54 & 13.81 & 13.04 &  1.22 \\
N581     &     Comparison &   20 53 34.8 &   47 55 54.7 & 12.69 & 11.97 & 11.55 & 11.18 &  0.72 \\
               & Reference  &   20 54 02.2 &   47 56 34.8 & 13.77 & 13.03 & 12.61 & 12.20 &  0.73 \\
\hline
\hline
\end{tabular}
\end{center}
\label{tab3}
\end{table*}


\section{New ephemeris}
Contact binary stars are known for having highly variable orbital periods. Accurate observation and measurement of the orbital period over time is required to help understand the dynamics of these systems and the effects that different factors have on their evolution. Additionally, the orbital period is a significant parameter whose relationship with other parameters of contact systems has always been investigated. This importance is much greater for the systems whose light curve analysis and changes in the orbital period have not yet been studied. These systems had few observations, so it is only possible to calculate new ephemeris for each of the systems.

Six of the systems did not have previously reported reliability minima. We also ignored the minimums reported with less than three decimal places. Therefore, for reference ephemerides, we used catalogs, databases, and studies (Table \ref{tab1} and Table \ref{tab4}).

We extracted 74 primary and secondary minima in $BVR_cI_c$ filters for the eight contact systems. All minima given in the Barycentric Julian Date and Barycentric Dynamical Time ($BJD_{TDB}$). Then, we averaged similar minima in different filters (Table \ref{tab4}). These minima were obtained by fitting the models to the observed light curves and existing minima using Gaussian and Cauchy distributions. So, we employed the MCMC method to determine the amount of uncertainty related to each value. The effect of starspots on light curves has been taken into account, however, it had little impact on the accuracy of minima extraction. Table \ref{tab4} contains the minima extracted in this study and collected from the literature. The epoch and O-C values of these minima were computed using the reference ephemeris. Also, a total of 411 minima were extracted from the TESS data for systems J069 and N581, which are presented in the appendix tables (A1 and A2). The new ephemeris for the systems is presented in Table \ref{tab5}.

\begin{figure*}
\begin{center}
\includegraphics[scale=0.20]{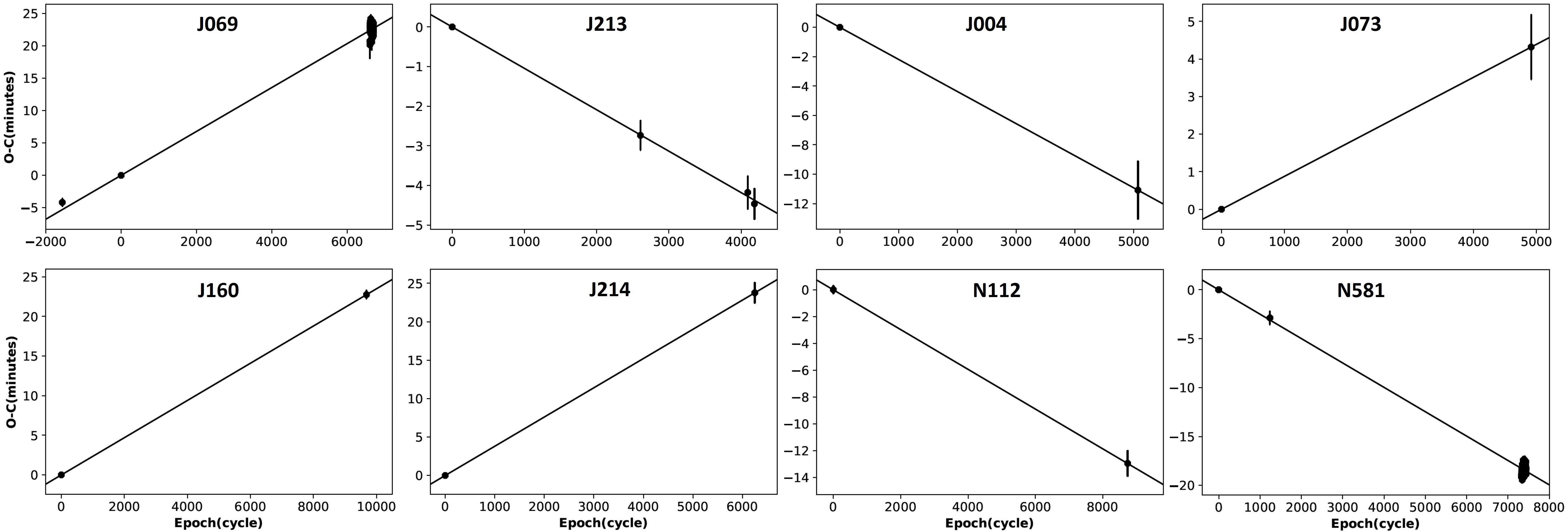}
    \caption{The O-C diagrams of our systems with a linear (black line) model.}
\label{Fig1}
\end{center}
\end{figure*}

Due to the short interval of observations for all studied systems, a linear fit was the best choice which we applied using least squares fitting. The O-C diagrams of the systems are presented in Figure \ref{Fig1}.

\begin{table*}
\caption{Extracted ground-based and gathered literature times of minima.}
\centering
\begin{center}
\footnotesize
\begin{tabular}{c c c c c c}
 \hline
 \hline
 System & Min.($BJD_{TDB}$) & Error & Epoch & O-C & Reference\\
\hline
J069	&	2457696.79502	&	0.00043	&	-1560	&	-0.0029	&	This study	\\
	&	2457696.93189	&	0.00033	&	-1559.5	&	-0.0016	&	This study	\\
	&	2458119.85553	&		&	0	&	0	&	ASAS-SN	\\
J213	&	2457323.73455	&		&	0	&	0	&	ASAS-SN	\\
	&	2457967.82090	&	0.00026	&	2606	&	-0.0019	&	This study	\\
	&	2458318.90452	&	0.00035	&	4026.5	&	-0.0033	&	This study	\\
	&	2458334.84642	&	0.00029	&	4091	&	-0.0029	&	This study	\\
	&	2458334.96958	&	0.00039	&	4091.5	&	-0.0033	&	This study	\\
	&	2458357.70734	&	0.00100	&	4183.5	&	-0.0039	&	This study	\\
	&	2458357.83174	&	0.00027	&	4184	&	-0.0031	&	This study	\\
J004	&	2458271.44264	&		&	0	&	0	&	ZTF	\\
	&	2459520.74918	&	0.00102	&	5071.5	&	-0.0058	&	This study	\\
	&	2459520.87037	&	0.00137	&	5072	&	-0.0077	&	This study	\\
J073	&	2456940.03906	&		&	0	&	0	&	ASAS-SN	\\
	&	2458103.75560	&	0.00095	&	4762.5	&	0.0016	&	This study	\\
	&	2458140.77604	&	0.00060	&	4914	&	0.0030	&	This study	\\
	&	2458894.71672	&	0.00103	&	7999.5	&	0.0030	&	This study	\\
J160	&	2457413.11089	&		&	0	&	0	&	ASAS-SN	\\
	&	2459682.84686	&	0.00043	&	9683.5	&	0.0146	&	This study	\\
	&	2459682.96524	&	0.00038	&	9684	&	0.0158	&	This study	\\
J214	&	2456835.98927	&		&	0	&	0	&	ASAS-SN	\\
	&	2458336.90181	&	0.00093	&	6249	&	0.0165	&	This study	\\
N112	&	2456144.36521	&	0.00021	&	0	&	0	&	\cite{2015RAA....15.1493K}	\\
	&	2458335.82384	&	0.00067	&	8740	&	-0.0090	&	This study	\\
	&	2458335.94837	&	0.00050	&	8740.5	&	-0.0098	&	This study	\\
N581	&	2458013.50209	&	0.00004	&	-4	&	0.0005	&	\cite{2020BlgAJ..32...71K}	\\
	&	2458014.36618	&	0.00005	&	-0.5	&	-0.0007	&	\cite{2020BlgAJ..32...71K}	\\
	&	2458014.49053	&	0.00011	&	0	&	0	&	\cite{2020BlgAJ..32...71K}	\\
	&	2458320.80225	&	0.00048	&	1239	&	-0.0020	&	This study	\\
	&	2458320.92456	&	0.00042	&	1239.5	&	-0.0033	&	This study	\\
 \hline
 \hline
\end{tabular}
\end{center}
\label{tab4}
\end{table*}

\begin{table*}
\caption{New ephemeris of the systems.}
\centering
\begin{center}
\footnotesize
\begin{tabular}{c c c}
 \hline
 \hline
 System & & New ephemeris\\
\hline
J069 & & Min.$I$($BJD_{TDB}$) = $2458119.85596(64)+0.271193109(98)\times E$\\
J213 & & Min.$I$($BJD_{TDB}$) = $2457323.73461(31)+0.247155087(86)\times E$\\
J004 & & Min.$I$($BJD_{TDB}$) = $2458271.44264(26)+0.246338470(338)\times E$\\
J073 & & Min.$I$($BJD_{TDB}$) = $2456940.03923(72)+0.244349995(137)\times E$\\
J160 & & Min.$I$($BJD_{TDB}$) = $2457413.11089(84)+0.234392168(106)\times E$\\
J214 & & Min.$I$($BJD_{TDB}$) = $2456835.98927(5)+0.240184436(9)\times E$\\
N112 & & Min.$I$($BJD_{TDB}$) = $2456144.36521(59)+ 0.250738926(83)\times E$\\
N581 & & Min.$I$($BJD_{TDB}$) = $2458014.49069(20)+0.247224846(28)\times E$\\
\hline
\hline
\end{tabular}
\end{center}
\label{tab5}
\end{table*}

\section{Light curve analysis}
The light curve analysis of the contact systems was performed using the Python code of PHOEBE version 2.4.9 (\citealt{2005ApJ...628..426P}, \citealt{2016ApJS..227...29P}, \citealt{2020ApJS..250...34C}). Six of the eight systems are investigated for the first time with the exception of N112 and N581. We adopted a contact mode because all eight systems displayed typical W UMa-type eclipsing binary light curves.

Accordingly, $g_1=g_2=0.32$ (\citealt{1967ZA.....65...89L}) and $A_1=A_2=0.5$ (\citealt{1969AcA....19..245R}) were assumed as the gravity-darkening coefficients and the bolometric albedo, respectively. The stellar atmosphere was modeled using the \cite{2004A&A...419..725C} study, and the limb darkening coefficients were included as a free parameter to the PHOEBE code.

We used the ($B-V$) color index value derived from our observations and calibration (\citealt{2000yCat.1259....0H}), along with the \cite{1996ApJ...469..355F} study's table, to calculate the primary star's temperature for the initial light curve solutions. We compared the measured temperature to those of Gaia DR2\footnote{\url{https://gea.esac.esa.int /archive/}} and the $P-T_1$ relation (\citealt{2022MNRAS.510.5315P}) that the values were close to each other. In this study, all of the initial temperatures were attributed to the hotter star after the morphology of the light curves was taken into consideration. In addition, we considered that the W UMa-type systems are known as Low-Temperature Contact Binaries (LTCBs), and the difference between the temperatures of two components in these types of eclipsing stars are close to zero (\citealt{2005ApJ...629.1055Y}). However, the temperature difference between the two stars in J004 is higher than in other systems.

\begin{figure*}
\begin{center}
\includegraphics[scale=0.32]{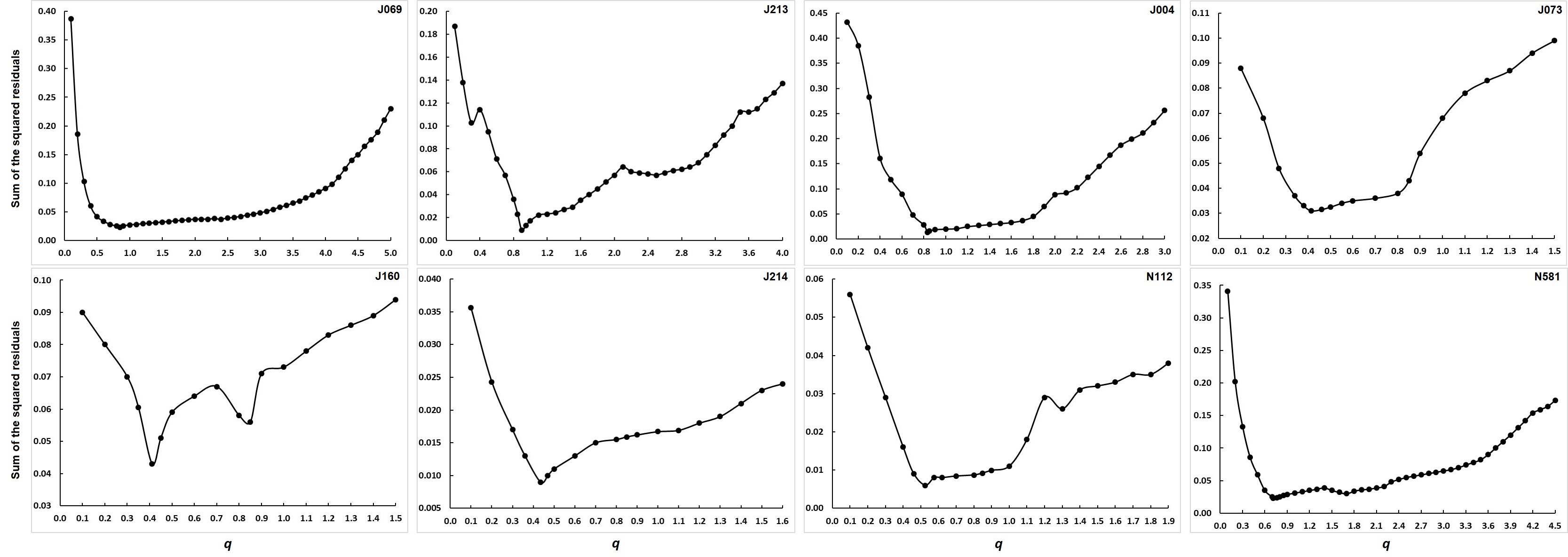}
    \caption{Sum of the squared residuals as a function of the mass ratio.}
\label{Fig2}
\end{center}
\end{figure*}

We carried out the required standards to obtain the mass ratio of the systems using the $q$-search method in the photometric observations. The results of $q$-search for eight systems are shown in Figure \ref{Fig2}. The $q$-search curves show a sharp bottom, making it easy to find a good mass ratio. Furthermore, it is evident from the curves in Figure \ref{Fig2} that all of the mass ratios found for the systems are smaller than 1.

The asymmetry in the brightness of maxima in the light curve of eclipsing binary stars is an indication of the well-known O'Connell effect (\citealt{1951PRCO....2...85O}). A possible explanation for this phenomenon could be the presence of magnetic activity on the surface of the star which causes the existence of starspot(s)(\citealt{2015NewA...37...64T}). Accordingly, six systems required a cold starspot in the light curve solution.

\begin{figure*}
\begin{center}
\includegraphics[scale=0.436]{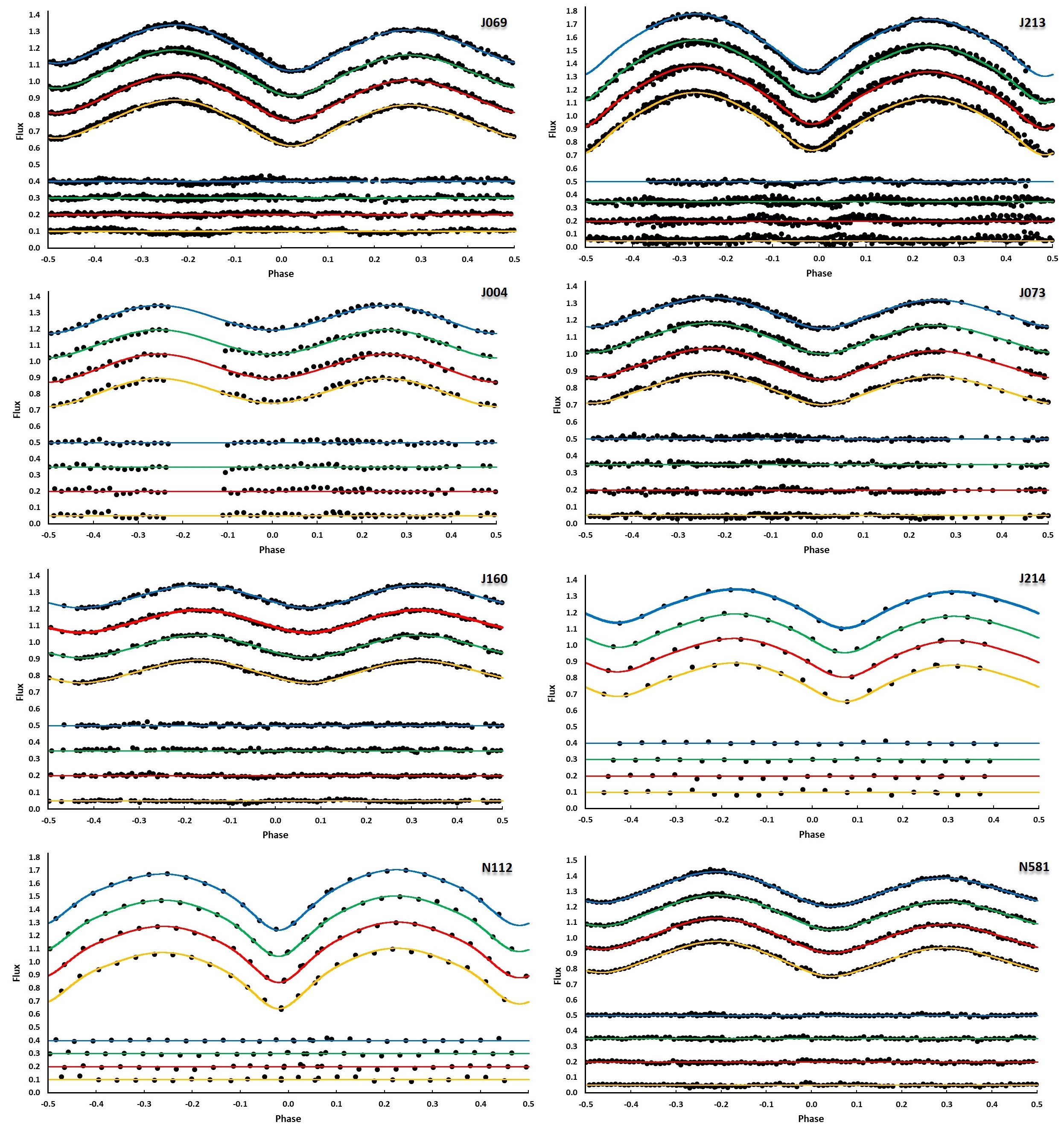}
    \caption{The systems' observed light curves are represented by the black dots, and the light curve solutions were used to produce the synthetic light curves. The relative flux was shifted freely while the orbital phase was kept constant. As indicated in the colors, the filters are $BVR_cI_c$ from top to bottom, respectively.}
\label{Fig3}
\end{center}
\end{figure*}

\begin{figure*}
\begin{center}
\includegraphics[scale=0.28]{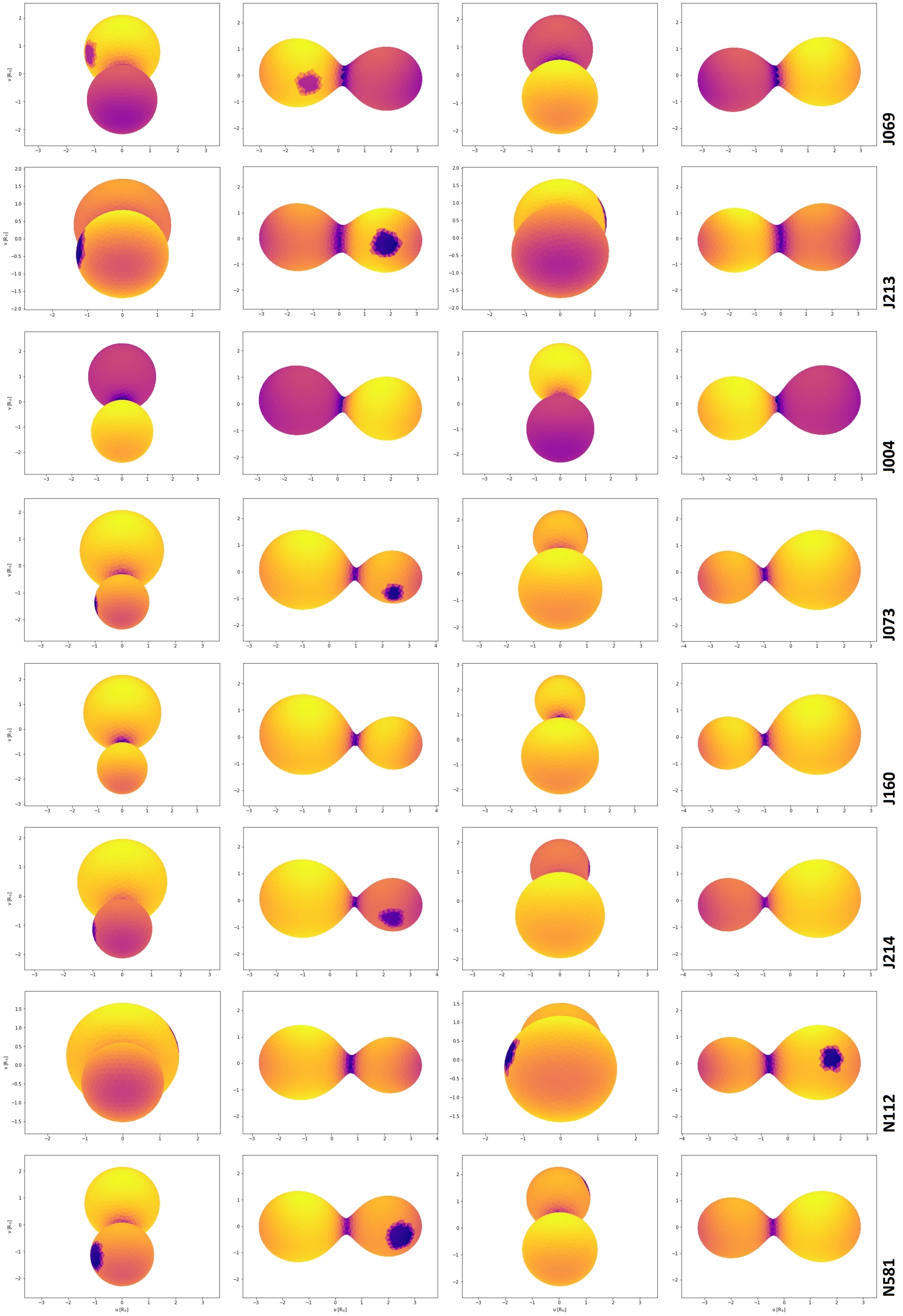}
    \caption{Three-dimensional view of the systems based on their light curve solutions.}
\label{Fig4}
\end{center}
\end{figure*}

Then, to improve the output of light curve solutions and obtain final results, we used PHOEBE's optimization tool. According to the residuals of the J213 and J069 systems, an additional trend can be seen; it seems necessary to pay attention to it in future studies. Figure \ref{Fig3} displays the observed and final synthetic light curves for $BVR_cI_c$ filters, and Table \ref{tab6} contains the results of the light curve solutions for each of the systems. Figure \ref{Fig4} obtained from the PHOEBE code and shows the systems in three dimensions, together with the starspots on the stars. The color in Figure \ref{Fig4} corresponds to the temperature differences on the star's surface.

\begin{table*}
\caption{Light curve solutions for the systems.}
\centering
\begin{center}
\footnotesize
\begin{tabular}{c c c c c c c c c}
 \hline
 \hline
Parameter & J069 & J213 & J004 & J073 & J160 & J214	& N112 & N581\\
\hline
$	T_1 (K)	$	&	4511(56)	&	4719(23)	&	4073(99)	&	4778(52)	&	4420(28)	&	4742(91)	&	4882(31)	&	4969(50)	\\
$	T_2 (K)	$	&	4189(81)	&	4839(44)	&	4575(87)	&	4675(39)	&	4378(25)	&	4464(87)	&	4785(27)	&	4835(34)	\\
$	q=M_2/M_1	$	&	0.850(23)	&	0.898(37)	&	0.833(30)	&	0.414(54)	&	0.411(49)	&	0.435(46)	&	0.525(30)	&	0.712(33)	\\
$	\Omega_1=\Omega_2	$	&	3.416(86)	&	3.411(99)	&	3.426(93)	&	2.682(55)	&	2.685(59)	&	2.735(73)	&	2.868(48)	&	3.214(51)	\\
$	f	$	&	0.182(16)	&	0.343(36)	&	0.102(8)	&	0.095(9)	&	0.063(6)	&	0.051(5)	&	0.179(17)	&	0.124(11)	\\
$	i^{\circ}	$	&	59.94(39)	&	75.97(52)	&	50.47(32)	&	55.94(64)	&	49.97(71)	&	61.81(43)	&	78.13(37)	&	56.19(69)	\\
$	l_1/l_{tot}	$	&	0.655(4)	&	0.483(2)	&	0.357(5)	&	0.720(2)	&	0.706(2)	&	0.764(4)	&	0.670(2)	&	0.627(3)	\\
$	l_2/l_{tot}	$	&	0.345(4)	&	0.517(2)	&	0.643(5)	&	0.280(2)	&	0.294(2)	&	0.236(4)	&	0.330(2)	&	0.373(3)	\\
$	r_{1(mean)}	$	&	0.411(3)	&	0.422(2)	&	0.405(3)	&	0.465(2)	&	0.464(2)	&	0.458(2)	&	0.451(2)	&	0.420(2)	\\
$	r_{2(mean)}	$	&	0.382(2)	&	0.403(2)	&	0.373(3)	&	0.312(2)	&	0.309(2)	&	0.313(2)	&	0.339(2)	&	0.361(3)	\\
	Phase shift		&	0.02(1)	&	-0.01(1)	&	0.00(1)	&	0.01(1)	&	0.07(1)	&	0.07(1)	&	-0.01(1)	&	0.04(1)	\\
\hline
	Col.(deg)		&	79(1)	&	98(1)	&		&	86(1)	&		&	87(1)	&	73(1)	&	101(1)	\\
	Long.(deg)		&	296(1)	&	100(1)	&		&	101(1)	&		&	98(1)	&	103(1)	&	116(1)	\\
	Rad.(deg)		&	19(1)	&	23(2)	&		&	19(1)	&		&	22(1)	&	16(1)	&	25(2)	\\
	$T_{spot}/T_{star}$		&	0.91(2)	&	0.90(2)	&		&	0.88(2)	&		&	0.91(2)	&	0.89(2)	&	0.88(2)	\\
	Component		&	Primary	&	Secondary	&		&	Secondary	&		&	Secondary	&	Primary	&	Secondary	\\
\hline
\hline
\end{tabular}
\end{center}
\label{tab6}
\end{table*}

\section{Absolute parameters}
\subsection{Estimation of values}
Estimating the absolute parameters is always one of the important goals of any study on contact binary systems. Some common methods require finding the primary star in the binary system based on the relationships between parameters such as period and primary star mass (e.g. \citealt{2003MNRAS.342.1260Q}).

In the absence of spectroscopic data, it is sometimes difficult to find a more massive star to perform calculations. Six of our target systems have a specific status based on the morphology of the light curves, and the more massive star can be introduced as the hotter star. On the other hand, using Gaia parallax is a good method to estimate the absolute parameters of the contact binary when we only have access to photometric data (\citealt{kjurkchieva2019w}, \citealt{2022MNRAS.510.5315P}). We used this method for our two systems where $T_2$ was hotter than $T_1$ to make sure which was the more massive component. It was not possible to use this method for all target systems.

There is a large parallax error for the J069 and N581 binary systems (Table \ref{tab1}). Systems with galactic coordinates b between +5 and -5 usually have large parallax errors, making it difficult to derive the extinction coefficient $A_V$ value with reasonable accuracy. Especially for the J069 system, $A_V$ is close to 1, and more interestingly, the Gaia DR3 parallax and temperature of this system show a big difference compared to Gaia DR2 results. 
The RUWE values from Gaia can give a better view, and this index should be in the maximum range of 1.4 (\citealt{lindegren2018re}). We have listed the RUWE of each system in Table \ref{tab1}, which shows that six systems were in the range, but systems J069 and N581 were much outside the range. Therefore, it can be concluded that for these two systems, the calculated values of parallax are not reliable, so the calculation of $A_V$ is also affected (\citealt{lindegren2018re}). It should be noted that the RUWE value for J069 and N581 might suggest some extra complexity, like the existence of a hidden third component that requires more future studies on these systems.

Fortunately, for J213 and J004, there were suitable conditions to estimate the masses of the stars with the Gaia DR3 parallax method and gain confidence. The results show that star 1 is more massive in the J213 and J004 systems.
\\
\\
We used the $P-M_1$ relation from the \cite{2022MNRAS.510.5315P} study. This relation is the result of estimating the absolute parameters of 118 systems using Gaia DR3's parallax method (Equation \ref{eq1}). It is used to estimate the mass of the more massive star in the system.

\begin{equation}\label{eq1}
M_1=(2.924\pm0.075)P+(0.147\pm0.029)
\end{equation}

Then, using the orbital period of each system and Kepler's third law the value of $a(R_{\odot})$ was calculated. According to the relationship between $a(R_{\odot})$ and $r_{mean}$, the radius of each star can be obtained. It is possible to determine each star's luminosity by knowing its temperature and radius. The bolometric absolute magnitude can be calculated by using Pogson’s famous relation (\citealt{1856MNRAS..17...12P}). It is possible to calculate the surface gravity of each star using the relationship between the mass and the radius. Finally, using the equation from the \cite{2006MNRAS.373.1483E} study, the orbital angular momentum ($J_0$) of the systems was computed (Equation \ref{eq2}):

\begin{equation}\label{eq2}
J_0=\frac{q}{(1+q)^2} \sqrt[3] {\frac{G^2}{2\pi}M^5P}
\end{equation}

where $q$ is the mass ratio, $M$ is the total mass of the system, $P$ is the orbital period, and $G$ is the gravitational constant. The units in equation \ref{eq2} are based on CGS.

The uncertainties of the absolute parameters were calculated considering the error bars of the associated parameters. The results of estimating the absolute parameters are shown in Table \ref{tab7}.

\begin{table*}
\caption{Estimated absolute parameters of the systems.}
\centering
\begin{center}
\footnotesize
\begin{tabular}{c c c c c c c c c}
 \hline
 \hline
Parameter & J069 & J213 & J004 & J073 & J160 & J214	& N112 & N581\\
\hline
$	M_1(M_\odot)	$	&	0.940(49)	&	0.870(48)	&	0.867(47)	&	0.861(47)	&	0.832(47)	&	0.849(47)	&	0.874(48)	&	0.870(48)	\\
$	M_2(M_\odot)	$	&	0.799(65)	&	0.781(77)	&	0.722(67)	&	0.357(69)	&	0.342(62)	&	0.369(62)	&	0.459(53)	&	0.619(64)	\\
$	R_1(R_\odot)	$	&	0.871(73)	&	0.826(69)	&	0.782(65)	&	0.817(51)	&	0.783(48)	&	0.795(50)	&	0.825(55)	&	0.795(61)	\\
$	R_2(R_\odot)	$	&	0.810(66)	&	0.789(66)	&	0.720(61)	&	0.548(35)	&	0.521(33)	&	0.544(35)	&	0.620(42)	&	0.683(55)	\\
$	L_1(L_\odot)	$	&	0.283(66)	&	0.305(60)	&	0.152(44)	&	0.313(56)	&	0.211(33)	&	0.288(63)	&	0.349(58)	&	0.347(71)	\\
$	L_2(L_\odot)	$	&	0.182(48)	&	0.308(67)	&	0.205(55)	&	0.129(22)	&	0.090(14)	&	0.106(24)	&	0.182(30)	&	0.230(46)	\\
$	M_{bol1}(mag)	$	&	6.099(227)	&	6.018(195)	&	6.779(279)	&	5.990(178)	&	6.419(158)	&	6.080(215)	&	5.874(168)	&	5.878(203)	\\
$	M_{bol2}(mag)	$	&	6.580(253)	&	6.009(214)	&	6.453(258)	&	6.951(172)	&	7.344(160)	&	7.169(221)	&	6.581(168)	&	6.325(198)	\\
$	log(g)_1(cgs)	$	&	4.531(47)	&	4.543(47)	&	4.590(47)	&	4.549(29)	&	4.571(28)	&	4.566(30)	&	4.546(33)	&	4.577(41)	\\
$	log(g)_2(cgs)	$	&	4.524(34)	&	4.536(29)	&	4.582(32)	&	4.513(22)	&	4.538(19)	&	4.535(12)	&	4.514(10)	&	4.561(24)	\\
$	a(R_\odot)	$	&	2.120(160)	&	1.958(154)	&	1.930(146)	&	1.756(101)	&	1.688(97)	&	1.737(101)	&	1.830(113)	&	1.893(135)	\\
$	logJ_0	$	&	51.70(5)	&	51.65(5)	&	51.62(5)	&	51.35(9)	&	51.32(8)	&	51.36(8)	&	51.45(6)	&	51.57(6)	\\
\hline
\hline
\end{tabular}
\end{center}
\label{tab7}
\end{table*}

\subsection{Comparison of estimated mass}
\cite{2022AJ....164..202L} presented a relationship between orbital period and semi-major axis ($a$) for contact binary systems. They used 168 systems with temperatures less than 10000 K.

\begin{figure*}
\begin{center}
\includegraphics[scale=0.48]{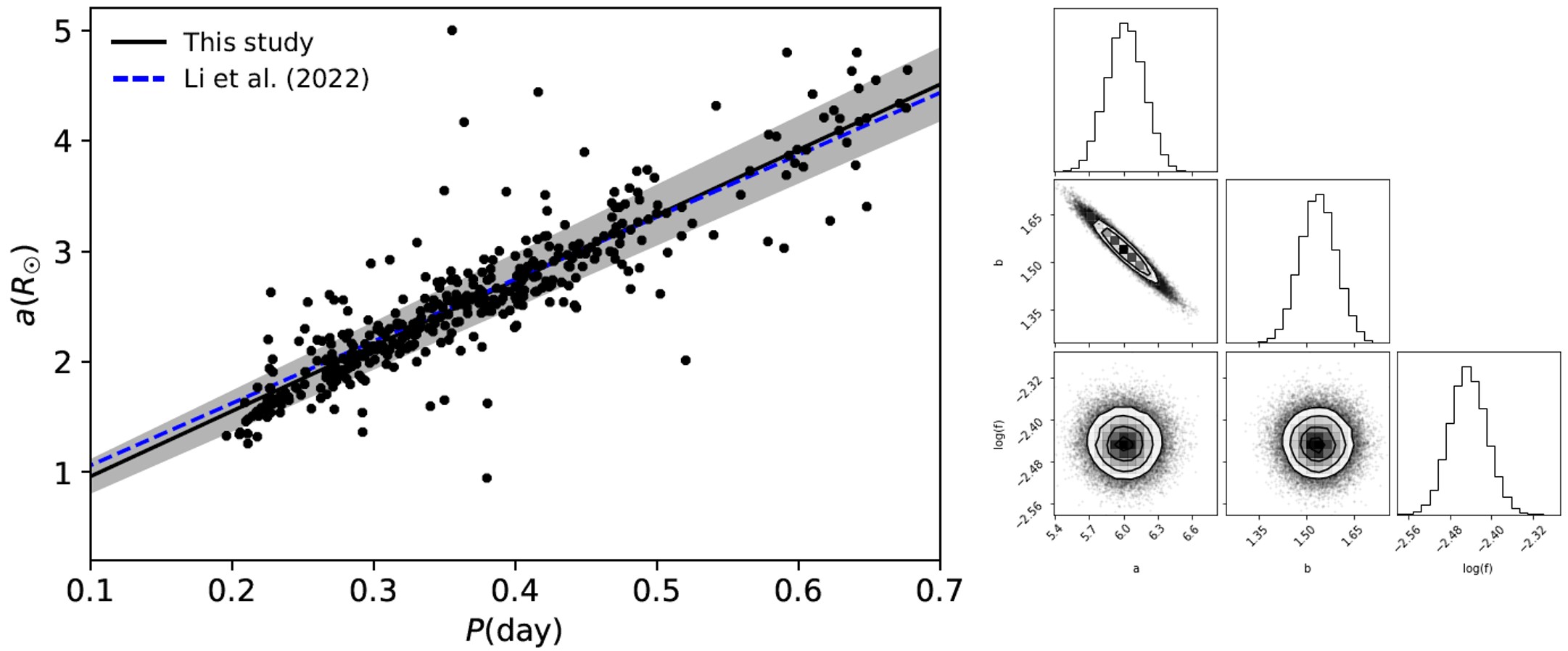}
    \caption{The relationship between the orbital period and the semi-major axis of contact systems, as well as the corner plot of the posterior distribution based on MCMC sampling. The selected contact systems have a maximum orbital period of 0.7 days.}
\label{Fig5}
\end{center}
\end{figure*}

Therefore, to examine the method we used to estimate the absolute parameters, we also calculated the mass of the stars for eight systems using the $P-a$ relationship. First, we updated the $P-a$ relationship using 414 contact systems with orbital periods less than 0.7 days taken from the \cite{2021ApJS..254...10L} study (Figure \ref{Fig5}). The updated empirical relationship between the orbital period and semi-major axis is presented in Equation \ref{eq3}:

\begin{equation}\label{eq3}
a=(0.372_{\rm-0.114}^{+0.113})+(5.914_{\rm-0.298}^{+0.272})\times P 
\end{equation}

Then, we calculated the mass and uncertainty of each star by using Kepler's well-known third law equation and the mass ratio from the light curve solutions.
\\
\\
In the \cite{2022MNRAS.510.5315P} study, a method for estimating $M_1$ using the ML method is proposed. We updated the presented model by adding 134 contact systems from the \cite{2021ApJS..254...10L} study whose spectroscopic observations were used for light curve analysis in the ML training process.

Considering the dual relationships between $P-M_1$, $P-T_1$ and $M_1-T_1$ (e.g. \citealt{2005ApJ...629.1055Y}, \citealt{2018PASJ...70...90K}, \citealt{2022MNRAS.510.5315P}), the relationship between $P-M_1-T_1$ parameters can be investigated. We used a three-layer deep ANN model and the Keras library on TensorFlow (\citealt{2017..Book..Brownlee}, \citealt{2016arXiv160304467A}). This model takes $P$ and $T_1$ as inputs and produces the output $M_1$. In our model, Mean Squared Error (MSE) is used for the loss function and Adaptive Movement Estimation (ADAM) is employed in optimization (\citealt{2014arXiv1412.6980K}). ADAM is an extension of gradient descent optimization, which has better performance for numerical output than other optimization methods. The rest of the work done is the same as the study of \cite{2022MNRAS.510.5315P}. By adding to the number of samples, our update shows a 4\% reduction in the amount of MSE.

\begin{figure*}
\begin{center}
\includegraphics[scale=0.19]{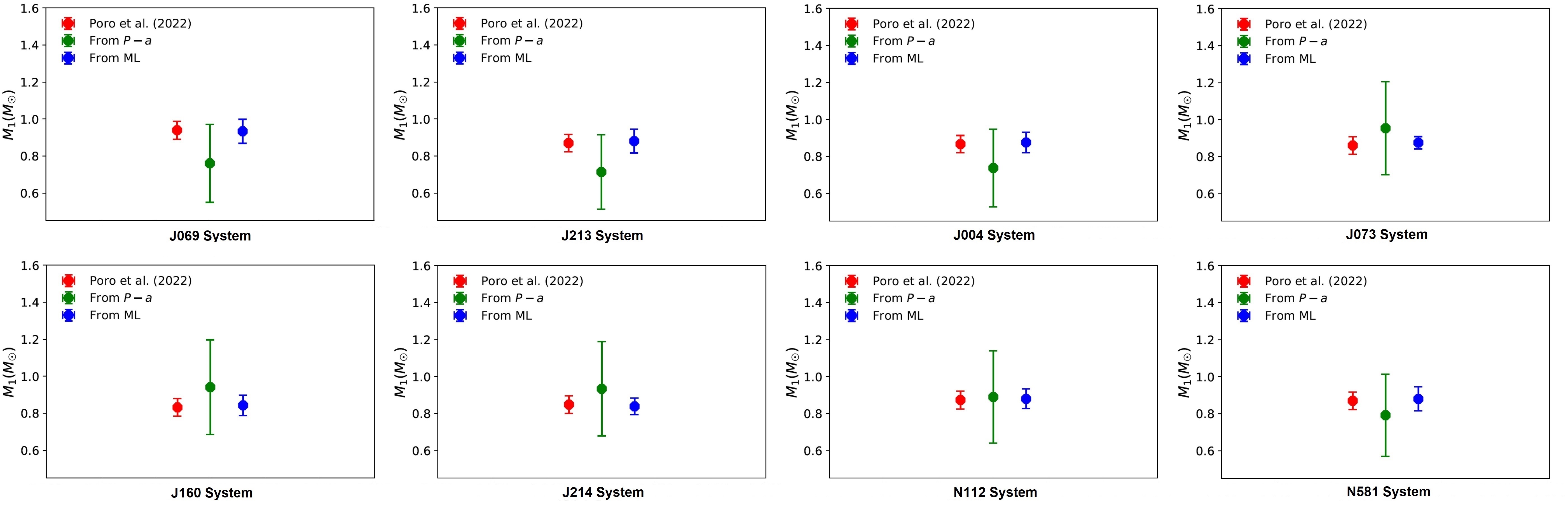}
    \caption{Comparison between the three methods mentioned in the text for estimating the mass of the primary star.}
\label{Fig6}
\end{center}
\end{figure*}

Therefore, the ML method was used to estimate $M_1$ in our systems using the $P$ and $T_1$ parameters. The results of all methods are presented in Figure \ref{Fig6}.

\section{Discussion and conclusion}
We presented a study of eight W UMa-type binary systems. The systems J069, J213, J004, J073, J160, and J214 which were analyzed for the first time. Observations of our systems were performed in 18 nights with multiband standard filters. According to our analysis, the following can be presented as conclusions:
\\
\\
A) We extracted minima from our light curves in all observational filters. Unfortunately, these systems have had few previous observations. So, the linear fit was the best choice in the O-C diagrams. Based on the reference ephemeris, we calculated and presented a new ephemeris for each system.
\\
\\
B) The temperature range of the stars in the systems is between 4100 K and 5000 K. The lowest temperature difference between two companion stars is related to the J160 system at a rate of 42 K, and the largest is for J004 at 502 K. According to the temperature of the stars and the \cite{2000asqu.book.....C} study, all of them are in the K spectral category. Specifically, the stars associated with system J069 are suggested as K5, J213 as K2; J004 as K5; J073 as K2; J160 as K5; J214 as K3 and K5, N112 as K2, N581 as K1 and K2 spectral types, respectively.

\begin{figure*}
\begin{center}
\includegraphics[scale=0.34]{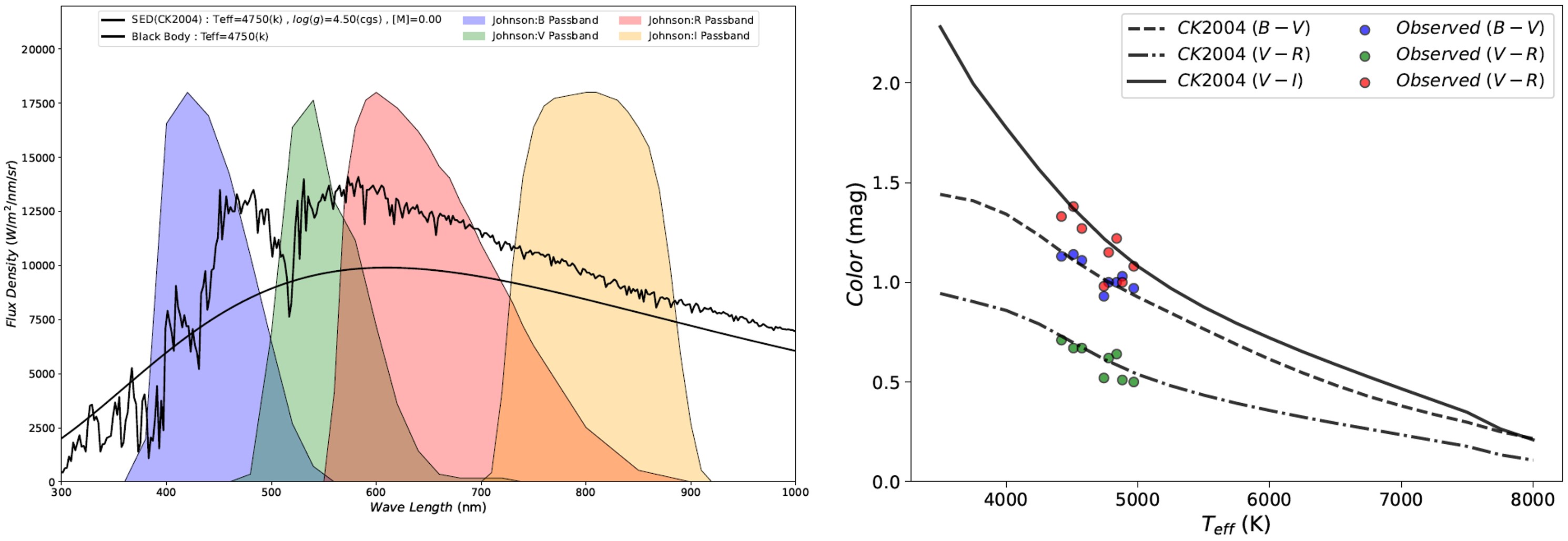}
    \caption{Left: The SED diagram for a sample star. Right: The Color-$T_{eff}$ diagram along with the position of the hotter stars.}
\label{fig7}
\end{center}
\end{figure*}

\begin{figure*}
\begin{center}
\includegraphics[scale=0.50]{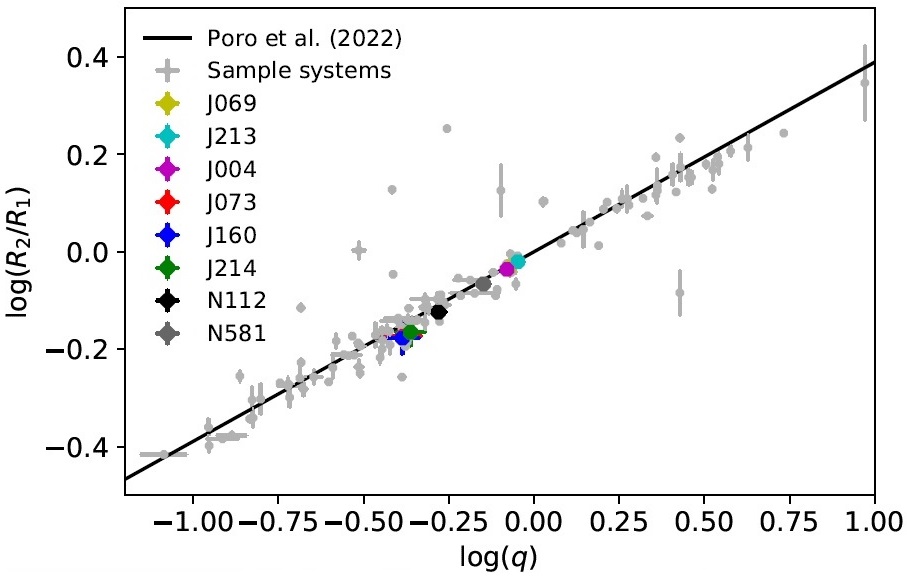}
    \caption{The mass ratio (from the light curve solutions) and radius ratio (from the absolute parameter estimation) relationship diagram for contact binary systems and the position of the eight studied systems on the relationship fit.}
\label{fig8}
\end{center}
\end{figure*}

We also made a comparison of the ($B-V$) obtained from our observations with the temperature results from the light curve analysis using a synthetic atmospheric model that is considered in this study. Based on the \cite{2004A&A...419..725C} atmosphere model, Figure \ref{fig7}(left) depicts the Spectral Energy Distribution (SED) diagram for the star with an effective temperature of 4750 K, $log(g)=4.50$, and metal abundance equal to 0. It also shows the passband transmission function of the $B$, $V$, $R$, and $I$ filters. The blackbody curve with an effective temperature of 4750 K is provided for comparison.
Furthermore, Figure \ref{fig7}(right) displays the ($B-V$), ($V-R$), and ($V-I$) plots according to effective temperature using the atmospheric model developed by \cite{2004A&A...419..725C}. The position of the stars in this study is indicated in the diagram.
\\
\\
C) Based on $q$-search and then light curve solutions, all eight systems have mass ratios of less than 1. It can be seen in Figure \ref{Fig2} that all $q$-search curves have a clear minimum sum of the squared residuals, and the mass ratios have been obtained with proper accuracy. According to the \cite{2022PASP..134f4201P} study the mass ratio has a significant relationship with the radius ratio in contact binary systems. Figure \ref{fig8} shows how the values obtained from the light curve solutions and the absolute parameters for $q$ and $R_{ratio}$ lie along the theoretical fit. Additionally, Figure \ref{fig8} includes 118 sample systems from the \cite{2022MNRAS.510.5315P} study.
\\
\\
D) We used the $P-M_1$ relationship from the \cite{2022MNRAS.510.5315P} study to estimate the absolute parameters. This equation estimates the mass of the more massive star in the system, and then the rest of the parameters can be estimated using the light curve solutions and the orbital periods (Table \ref{tab7}).

We employed additional methods to evaluate the $M_1$ estimations. So, we updated the $P-a$ empirical relationship with a sample of 414 contact systems. It should be noted that using the $P-a$ relationship for calculating the mass of the stars gives a large uncertainty.

Another method to evaluate was the use of machine learning, which was utilized in the \cite{2022MNRAS.510.5315P} study. So, we updated the relationship between $P-T_1-M_1$ using the ANN method. For this purpose, we added 134 contact systems from the \cite{2021ApJS..254...10L} study that were observed and analyzed using spectroscopic data to the previous sample.

\begin{figure*}
\begin{center}
\includegraphics[scale=0.36]{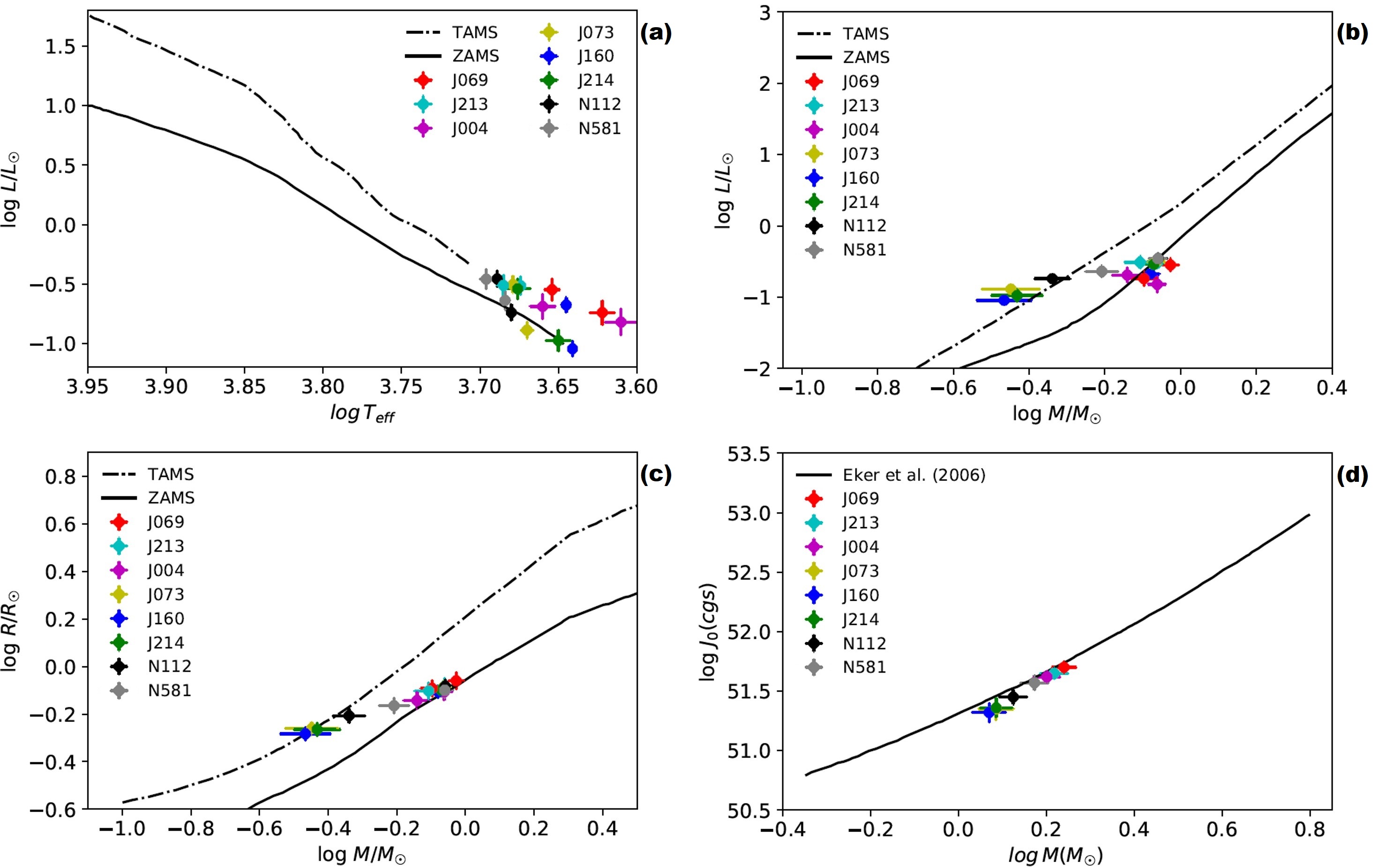}
    \caption{The HR, $logM-logL$, $logM-logR$, and $logJ_0-logM_{tot}$ diagrams for the systems.}
\label{fig9}
\end{center}
\end{figure*}

The results from the $P-a$ and ANN methods showed that the calculated masses were in good consistency with the \cite{2022MNRAS.510.5315P} method (Figure \ref{Fig6}).
\\
\\
E) We displayed the state of evolution of the systems on the Hertzsprung-Russell (HR), Mass-Luminosity ($M-L$), and Mass-Radius ($M-R$) diagrams based on the light curve analysis and the estimate of absolute parameters (Figure \ref{fig9}a,b,c). In each of these three diagrams, the position of companion stars was displayed relative to the theoretical Zero-Age Main Sequence (ZAMS) and Terminal-Age Main Sequence (TAMS) lines.

We estimated orbital angular momentum for each system (Table \ref{tab7}); and their locations are shown on the $logJ_0-logM$ diagram (Figure \ref{fig9}d). We used the parabolic line from the \cite{2006MNRAS.373.1483E} study that shows all our systems in the contact binary region.

The primary stars of the four systems J073, J160, J214, and N112 in the three HR, $M-L$, and $M-R$ diagrams were on or above the TAMS line, which shows that they were more evolved than secondary stars. These four binaries (J073, J160, J214, and N112) have the smallest $q$ and greater $log(g)$ difference between two stars than other systems.
According to Figure \ref{fig9}a,b,c diagrams, the stars of the J069, J213, J004, and N581 systems have almost the same evolutionary status.

\begin{figure*}
\begin{center}
\includegraphics[scale=0.43]{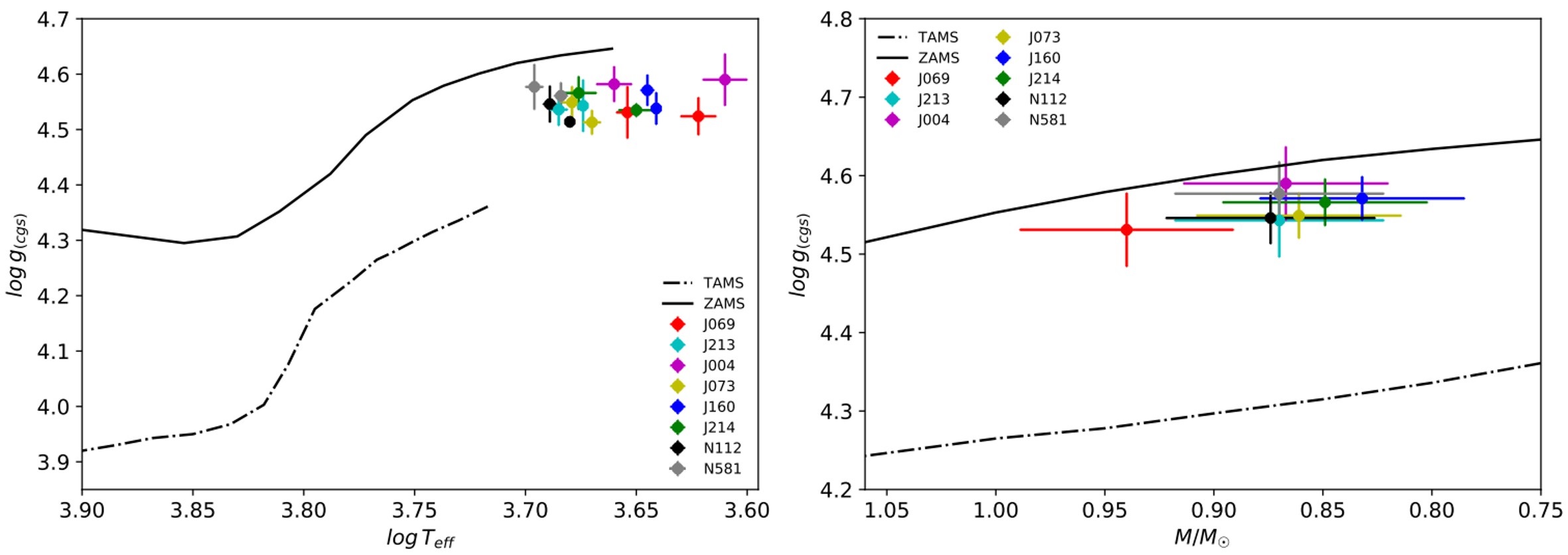}
    \caption{The $log(g)-logT$, and $log(g)-M$ diagrams.}
\label{fig10}
\end{center}
\end{figure*}

Additionally, we showed the positions of the stars on the $log(g)-logT$ and $log(g)-M$ diagrams (\ref{fig10}. Both diagrams came from the \cite{2015RAA....15.2244Y} study. We only showed the primary stars on the $log(g)-M$ diagram, since the \cite{2015RAA....15.2244Y} study provided the diagram based on the mass range of 0.75 to 10 $M_{\odot}$. The stars are situated between ZAMS and TAMS in both figures.
\\
\\
F) The light curve for the N112 system has already been studied. \cite{2015RAA....15.1493K} have obtained the following results using photometric observations in $V$ and $I_c$ filters from the light curves for this system: $q=0.986(6)$, $i=74.36(26)$, $T_1=4600(90)$, $T_2=3969(70)$, $f=0.209$. This study did not estimate the absolute parameters of the system. It seems that the temperature difference between the two stars is 631 K, compared to 97 K in this study, and this is a significant difference. The mass ratio measured in the two studies shows a difference of 0.46, although both values are below $q=1$.

N581 had also been studied by \cite{2020BlgAJ..32...71K}. They presented the following results: $q=1.499(3)$, $i=51.7(1)$, $T_1=5596(27)$, $T_2=4811(29)$, $f=0.213$. The temperature difference between components in that study is equal to 785 K, which is a large value for contact systems. The most important difference between \cite{2020BlgAJ..32...71K} and this study is in the mass ratio values. In our study, $q=0.712(33)$ has been determined. Looking at the $q$-search curve in the study of \cite{2020BlgAJ..32...71K}, we find that the minimum extracted is not completely clear and the bottom of the curve is very close to each other in a large interval.
\\
\\
G) According to the analysis of light curves, all the targets were detected as contact binary systems.

\vspace{1cm}

\section*{Acknowledgements}
This manuscript has been prepared based on a multilateral collaboration between the BSN project (\url{http://bsnp.info}), the Raderon AI Lab (\url{https://raderonlab.ca}), and Erciyes University (\url{https://www.erciyes.edu.tr}). "This study was supported by the Scientific Research Projects Coordination Unit of Erciyes University (project number FBA-2022-11737)". Also, this work is based upon observations carried out at the Observatorio Astronómico Nacional on the Sierra San Pedro Mártir (OAN-SPM), Baja California, México. We made use of the SIMBAD database, which is operated by CDS in Strasbourg, France (\url{http://simbad.u-strasbg.fr/simbad/}). Thanks to Mehdi Khodadadilori for helping with the machine learning tests. We appreciate Jabar Rahimi's assistance in extracting the times of minima. Thanks as well to Paul D. Maley for making editorial corrections to the text. We express our gratitude to Edwin Budding for providing scientific and effective comments.


\section*{Data Availability}
The data are available as a machine-readable table in the online version of this paper.


\section*{ORCID iDs}
\noindent Atila Poro: 0000-0002-0196-9732\\
Mehmet Tanriver: 0000-0002-3263-9680\\
Raul Michel: 0000-0003-1263-808X\\
Ehsan Paki: 0000-0001-9746-2284\\

\clearpage
\appendix
\section*{Appendix: TESS Minima Times}
TESS data are available for two systems: J069 and N581, both with an exposure time of 200 seconds. Sectors 59 and 56 included data from systems J069 and N581, respectively. So, the times of minima that we extracted from these data are listed in the appendix tables.

\bsp


\bibliographystyle{mnras}
\bibliography{references}

\begin{thebibliography}{}
\makeatletter
\relax
\def\mn@urlcharsother{\let\do\@makeother \do\$\do\&\do\#\do\^\do\_\do\%\do\~}
\def\mn@doi{\begingroup\mn@urlcharsother \@ifnextchar [ {\mn@doi@}
  {\mn@doi@[]}}
\def\mn@doi@[#1]#2{\def\@tempa{#1}\ifx\@tempa\@empty \href
  {http://dx.doi.org/#2} {doi:#2}\else \href {http://dx.doi.org/#2} {#1}\fi
  \endgroup}
\def\mn@eprint#1#2{\mn@eprint@#1:#2::\@nil}
\def\mn@eprint@arXiv#1{\href {http://arxiv.org/abs/#1} {{\tt arXiv:#1}}}
\def\mn@eprint@dblp#1{\href {http://dblp.uni-trier.de/rec/bibtex/#1.xml}
  {dblp:#1}}
\def\mn@eprint@#1:#2:#3:#4\@nil{\def\@tempa {#1}\def\@tempb {#2}\def\@tempc
  {#3}\ifx \@tempc \@empty \let \@tempc \@tempb \let \@tempb \@tempa \fi \ifx
  \@tempb \@empty \def\@tempb {arXiv}\fi \@ifundefined
  {mn@eprint@\@tempb}{\@tempb:\@tempc}{\expandafter \expandafter \csname
  mn@eprint@\@tempb\endcsname \expandafter{\@tempc}}}

\bibitem[\protect\citeauthoryear{{Abadi} et~al.,}{{Abadi}
  et~al.}{2016}]{2016arXiv160304467A}
{Abadi} M.,  et~al., 2016, arXiv e-prints, \href
  {https://ui.adsabs.harvard.edu/abs/2016arXiv160304467A} {p. arXiv:1603.04467}

\bibitem[\protect\citeauthoryear{{Astropy Collaboration} et~al.,}{{Astropy
  Collaboration} et~al.}{2013}]{2013A&A...558A..33A}
{Astropy Collaboration} et~al., 2013, \mn@doi [\aap]
  {10.1051/0004-6361/201322068}, \href
  {https://ui.adsabs.harvard.edu/abs/2013A&A...558A..33A} {558, A33}

\bibitem[\protect\citeauthoryear{{Bradstreet} \& {Guinan}}{{Bradstreet} \&
  {Guinan}}{1994}]{1994ASPC...56..228B}
{Bradstreet} D.~H.,  {Guinan} E.~F.,  1994, in {Shafter} A.~W.,  ed.,
  Astronomical Society of the Pacific Conference Series Vol. 56, Interacting
  Binary Stars. p.~228

\bibitem[\protect\citeauthoryear{{Brownlee}}{{Brownlee}}{2016}]{2017..Book..Brownlee}
{Brownlee} J.,  2016, Machine Learning Mastery

\bibitem[\protect\citeauthoryear{{Castelli} \& {Kurucz}}{{Castelli} \&
  {Kurucz}}{2004}]{2004A&A...419..725C}
{Castelli} F.,  {Kurucz} R.~L.,  2004, \mn@doi [\aap]
  {10.1051/0004-6361:20040079}, \href
  {https://ui.adsabs.harvard.edu/abs/2004A&A...419..725C} {419, 725}

\bibitem[\protect\citeauthoryear{{Collins}, {Kielkopf}, {Stassun}  \&
  {Hessman}}{{Collins} et~al.}{2017}]{2017AJ....153...77C}
{Collins} K.~A.,  {Kielkopf} J.~F.,  {Stassun} K.~G.,   {Hessman} F.~V.,  2017,
  \mn@doi [\aj] {10.3847/1538-3881/153/2/77}, \href
  {https://ui.adsabs.harvard.edu/abs/2017AJ....153...77C} {153, 77}

\bibitem[\protect\citeauthoryear{{Conroy} et~al.,}{{Conroy}
  et~al.}{2020}]{2020ApJS..250...34C}
{Conroy} K.~E.,  et~al., 2020, \mn@doi [\apjs] {10.3847/1538-4365/abb4e2},
  \href {https://ui.adsabs.harvard.edu/abs/2020ApJS..250...34C} {250, 34}

\bibitem[\protect\citeauthoryear{{Cox}}{{Cox}}{2000}]{2000asqu.book.....C}
{Cox} A.~N.,  2000, {Allen's astrophysical quantities}

\bibitem[\protect\citeauthoryear{{Drake} et~al.,}{{Drake}
  et~al.}{2014a}]{2014ApJS..213....9D}
{Drake} A.~J.,  et~al., 2014a, \mn@doi [\apjs] {10.1088/0067-0049/213/1/9},
  \href {https://ui.adsabs.harvard.edu/abs/2014ApJS..213....9D} {213, 9}

\bibitem[\protect\citeauthoryear{{Drake} et~al.,}{{Drake}
  et~al.}{2014b}]{2014ApJ...790..157D}
{Drake} A.~J.,  et~al., 2014b, \mn@doi [\apj] {10.1088/0004-637X/790/2/157},
  \href {https://ui.adsabs.harvard.edu/abs/2014ApJ...790..157D} {790, 157}

\bibitem[\protect\citeauthoryear{{Eggleton}}{{Eggleton}}{2012}]{2012JASS...29..145E}
{Eggleton} P.~P.,  2012, \mn@doi [Journal of Astronomy and Space Sciences]
  {10.5140/JASS.2012.29.2.145}, \href
  {https://ui.adsabs.harvard.edu/abs/2012JASS...29..145E} {29, 145}

\bibitem[\protect\citeauthoryear{{Eker}, {Demircan}, {Bilir}  \&
  {Karata{\c{s}}}}{{Eker} et~al.}{2006}]{2006MNRAS.373.1483E}
{Eker} Z.,  {Demircan} O.,  {Bilir} S.,   {Karata{\c{s}}} Y.,  2006, \mn@doi
  [\mnras] {10.1111/j.1365-2966.2006.11073.x}, \href
  {https://ui.adsabs.harvard.edu/abs/2006MNRAS.373.1483E} {373, 1483}

\bibitem[\protect\citeauthoryear{{Eker}, {Demircan}  \& {Bilir}}{{Eker}
  et~al.}{2008}]{2008MNRAS.386.1756E}
{Eker} Z.,  {Demircan} O.,   {Bilir} S.,  2008, \mn@doi [\mnras]
  {10.1111/j.1365-2966.2008.13155.x}, \href
  {https://ui.adsabs.harvard.edu/abs/2008MNRAS.386.1756E} {386, 1756}

\bibitem[\protect\citeauthoryear{{Flower}}{{Flower}}{1996}]{1996ApJ...469..355F}
{Flower} P.~J.,  1996, \mn@doi [\apj] {10.1086/177785}, \href
  {https://ui.adsabs.harvard.edu/abs/1996ApJ...469..355F} {469, 355}

\bibitem[\protect\citeauthoryear{{Gaia Collaboration} et~al.,}{{Gaia
  Collaboration} et~al.}{2023}]{2023A&A...674A..33G}
{Gaia Collaboration} et~al., 2023, \mn@doi [\aap]
  {10.1051/0004-6361/202243709}, \href
  {https://ui.adsabs.harvard.edu/abs/2023A&A...674A..33G} {674, A33}

\bibitem[\protect\citeauthoryear{{Heinze} et~al.,}{{Heinze}
  et~al.}{2018}]{2018AJ....156..241H}
{Heinze} A.~N.,  et~al., 2018, \mn@doi [\aj] {10.3847/1538-3881/aae47f}, \href
  {https://ui.adsabs.harvard.edu/abs/2018AJ....156..241H} {156, 241}

\bibitem[\protect\citeauthoryear{{Hiltner}}{{Hiltner}}{1962}]{1962aste.book.....H}
{Hiltner} W.~A.,  1962, {Astronomical techniques.}

\bibitem[\protect\citeauthoryear{{Hoffman}, {Harrison}  \&
  {McNamara}}{{Hoffman} et~al.}{2009}]{2009AJ....138..466H}
{Hoffman} D.~I.,  {Harrison} T.~E.,   {McNamara} B.~J.,  2009, \mn@doi [\aj]
  {10.1088/0004-6256/138/2/466}, \href
  {https://ui.adsabs.harvard.edu/abs/2009AJ....138..466H} {138, 466}

\bibitem[\protect\citeauthoryear{{Hog} et~al.,}{{Hog}
  et~al.}{2000}]{2000yCat.1259....0H}
{Hog} E.,  et~al., 2000, VizieR Online Data Catalog, \href
  {https://ui.adsabs.harvard.edu/abs/2000yCat.1259....0H} {p. I/259}

\bibitem[\protect\citeauthoryear{{Kingma} \& {Ba}}{{Kingma} \&
  {Ba}}{2014}]{2014arXiv1412.6980K}
{Kingma} D.~P.,  {Ba} J.,  2014, \mn@doi [arXiv e-prints]
  {10.48550/arXiv.1412.6980}, \href
  {https://ui.adsabs.harvard.edu/abs/2014arXiv1412.6980K} {p. arXiv:1412.6980}

\bibitem[\protect\citeauthoryear{{Kjurkchieva}, {Dimitrov}  \&
  {Ibryamov}}{{Kjurkchieva} et~al.}{2015}]{2015RAA....15.1493K}
{Kjurkchieva} D.~P.,  {Dimitrov} D.~P.,   {Ibryamov} S.~I.,  2015, \mn@doi
  [Research in Astronomy and Astrophysics] {10.1088/1674-4527/15/9/006}, \href
  {https://ui.adsabs.harvard.edu/abs/2015RAA....15.1493K} {15, 1493}

\bibitem[\protect\citeauthoryear{Kjurkchieva, Popov, Eneva  \&
  Petrov}{Kjurkchieva et~al.}{2019}]{kjurkchieva2019w}
Kjurkchieva D.~P.,  Popov V.~A.,  Eneva Y.,   Petrov N.~I.,  2019, Research in
  Astronomy and Astrophysics, 19, 014

\bibitem[\protect\citeauthoryear{{Kjurkchieva}, {Popov}, {Eneva}  \&
  {Petrov}}{{Kjurkchieva} et~al.}{2020}]{2020BlgAJ..32...71K}
{Kjurkchieva} D.,  {Popov} V.,  {Eneva} Y.,   {Petrov} N.,  2020, Bulgarian
  Astronomical Journal, \href
  {https://ui.adsabs.harvard.edu/abs/2020BlgAJ..32...71K} {32, 71}

\bibitem[\protect\citeauthoryear{{Kouzuma}}{{Kouzuma}}{2018}]{2018PASJ...70...90K}
{Kouzuma} S.,  2018, \mn@doi [\pasj] {10.1093/pasj/psy086}, \href
  {https://ui.adsabs.harvard.edu/abs/2018PASJ...70...90K} {70, 90}

\bibitem[\protect\citeauthoryear{{Latkovi{\'c}}, {{\v{C}}eki}  \&
  {Lazarevi{\'c}}}{{Latkovi{\'c}} et~al.}{2021}]{2021ApJS..254...10L}
{Latkovi{\'c}} O.,  {{\v{C}}eki} A.,   {Lazarevi{\'c}} S.,  2021, \mn@doi
  [\apjs] {10.3847/1538-4365/abeb23}, \href
  {https://ui.adsabs.harvard.edu/abs/2021ApJS..254...10L} {254, 10}

\bibitem[\protect\citeauthoryear{{Li}, {Zhang}, {Han}  \& {Jiang}}{{Li}
  et~al.}{2007}]{2007ApJ...662..596L}
{Li} L.,  {Zhang} F.,  {Han} Z.,   {Jiang} D.,  2007, \mn@doi [\apj]
  {10.1086/517909}, \href
  {https://ui.adsabs.harvard.edu/abs/2007ApJ...662..596L} {662, 596}

\bibitem[\protect\citeauthoryear{{Li}, {Gao}, {Liu}, {Gao}, {Li}, {Chen}  \&
  {Sun}}{{Li} et~al.}{2022}]{2022AJ....164..202L}
{Li} K.,  {Gao} X.,  {Liu} X.-Y.,  {Gao} X.,  {Li} L.-Z.,  {Chen} X.,   {Sun}
  G.-Y.,  2022, \mn@doi [\aj] {10.3847/1538-3881/ac8ff2}, \href
  {https://ui.adsabs.harvard.edu/abs/2022AJ....164..202L} {164, 202}

\bibitem[\protect\citeauthoryear{Lindegren et~al.}{Lindegren
  et~al.}{2018}]{lindegren2018re}
Lindegren L.,  et~al., 2018, Gaia Technical Note: GAIA-C3-TN-LU-LL-124-01

\bibitem[\protect\citeauthoryear{{Lucy}}{{Lucy}}{1967}]{1967ZA.....65...89L}
{Lucy} L.~B.,  1967, \zap, \href
  {https://ui.adsabs.harvard.edu/abs/1967ZA.....65...89L} {65, 89}

\bibitem[\protect\citeauthoryear{{Nefs} et~al.,}{{Nefs}
  et~al.}{2012}]{2012MNRAS.425..950N}
{Nefs} S.~V.,  et~al., 2012, \mn@doi [\mnras]
  {10.1111/j.1365-2966.2012.21338.x}, \href
  {https://ui.adsabs.harvard.edu/abs/2012MNRAS.425..950N} {425, 950}

\bibitem[\protect\citeauthoryear{{O'Connell}}{{O'Connell}}{1951}]{1951PRCO....2...85O}
{O'Connell} D.~J.~K.,  1951, Publications of the Riverview College Observatory,
  \href {https://ui.adsabs.harvard.edu/abs/1951PRCO....2...85O} {2, 85}

\bibitem[\protect\citeauthoryear{{Pogson}}{{Pogson}}{1856}]{1856MNRAS..17...12P}
{Pogson} N.,  1856, \mn@doi [\mnras] {10.1093/mnras/17.1.12}, \href
  {https://ui.adsabs.harvard.edu/abs/1856MNRAS..17...12P} {17, 12}

\bibitem[\protect\citeauthoryear{{Poro} et~al.,}{{Poro}
  et~al.}{2022a}]{2022PASP..134f4201P}
{Poro} A.,  et~al., 2022a, \mn@doi [\pasp] {10.1088/1538-3873/ac71cd}, \href
  {https://ui.adsabs.harvard.edu/abs/2022PASP..134f4201P} {134, 064201}

\bibitem[\protect\citeauthoryear{{Poro} et~al.,}{{Poro}
  et~al.}{2022b}]{2022MNRAS.510.5315P}
{Poro} A.,  et~al., 2022b, \mn@doi [\mnras] {10.1093/mnras/stab3775}, \href
  {https://ui.adsabs.harvard.edu/abs/2022MNRAS.510.5315P} {510, 5315}

\bibitem[\protect\citeauthoryear{{Poro} et~al.,}{{Poro}
  et~al.}{2024}]{2024RAA....24a5002P}
{Poro} A.,  et~al., 2024, \mn@doi [Research in Astronomy and Astrophysics]
  {10.1088/1674-4527/ad0866}, \href
  {https://ui.adsabs.harvard.edu/abs/2024RAA....24a5002P} {24, 015002}

\bibitem[\protect\citeauthoryear{{Pr{\v{s}}a} \& {Zwitter}}{{Pr{\v{s}}a} \&
  {Zwitter}}{2005}]{2005ApJ...628..426P}
{Pr{\v{s}}a} A.,  {Zwitter} T.,  2005, \mn@doi [\apj] {10.1086/430591}, \href
  {https://ui.adsabs.harvard.edu/abs/2005ApJ...628..426P} {628, 426}

\bibitem[\protect\citeauthoryear{{Pr{\v{s}}a} et~al.,}{{Pr{\v{s}}a}
  et~al.}{2016}]{2016ApJS..227...29P}
{Pr{\v{s}}a} A.,  et~al., 2016, \mn@doi [\apjs] {10.3847/1538-4365/227/2/29},
  \href {https://ui.adsabs.harvard.edu/abs/2016ApJS..227...29P} {227, 29}

\bibitem[\protect\citeauthoryear{{Qian}}{{Qian}}{2003}]{2003MNRAS.342.1260Q}
{Qian} S.,  2003, \mn@doi [\mnras] {10.1046/j.1365-8711.2003.06627.x}, \href
  {https://ui.adsabs.harvard.edu/abs/2003MNRAS.342.1260Q} {342, 1260}

\bibitem[\protect\citeauthoryear{{Ruci{\'n}ski}}{{Ruci{\'n}ski}}{1969}]{1969AcA....19..245R}
{Ruci{\'n}ski} S.~M.,  1969, \actaa, \href
  {https://ui.adsabs.harvard.edu/abs/1969AcA....19..245R} {19, 245}

\bibitem[\protect\citeauthoryear{{Tavakkoli}, {Hasanzadeh}  \&
  {Poro}}{{Tavakkoli} et~al.}{2015}]{2015NewA...37...64T}
{Tavakkoli} F.,  {Hasanzadeh} A.,   {Poro} A.,  2015, \mn@doi [\na]
  {10.1016/j.newast.2014.12.004}, \href
  {https://ui.adsabs.harvard.edu/abs/2015NewA...37...64T} {37, 64}

\bibitem[\protect\citeauthoryear{{Terrell}, {Gross}  \& {Cooney}}{{Terrell}
  et~al.}{2012}]{2012AJ....143...99T}
{Terrell} D.,  {Gross} J.,   {Cooney} W.~R.,  2012, \mn@doi [\aj]
  {10.1088/0004-6256/143/4/99}, \href
  {https://ui.adsabs.harvard.edu/abs/2012AJ....143...99T} {143, 99}

\bibitem[\protect\citeauthoryear{{Tody}}{{Tody}}{1986}]{1986SPIE..627..733T}
{Tody} D.,  1986, in {Crawford} D.~L.,  ed.,  Society of Photo-Optical
  Instrumentation Engineers (SPIE) Conference Series Vol. 627, Instrumentation
  in astronomy VI. p.~733, \mn@doi{10.1117/12.968154}

\bibitem[\protect\citeauthoryear{{Torrealba} et~al.,}{{Torrealba}
  et~al.}{2015}]{2015MNRAS.446.2251T}
{Torrealba} G.,  et~al., 2015, \mn@doi [\mnras] {10.1093/mnras/stu2274}, \href
  {https://ui.adsabs.harvard.edu/abs/2015MNRAS.446.2251T} {446, 2251}

\bibitem[\protect\citeauthoryear{{Yakut} \& {Eggleton}}{{Yakut} \&
  {Eggleton}}{2005}]{2005ApJ...629.1055Y}
{Yakut} K.,  {Eggleton} P.~P.,  2005, \mn@doi [\apj] {10.1086/431300}, \href
  {https://ui.adsabs.harvard.edu/abs/2005ApJ...629.1055Y} {629, 1055}

\bibitem[\protect\citeauthoryear{{Y{\i}ld{\i}z}}{{Y{\i}ld{\i}z}}{2015}]{2015RAA....15.2244Y}
{Y{\i}ld{\i}z} M.,  2015, \mn@doi [Research in Astronomy and Astrophysics]
  {10.1088/1674-4527/15/12/012}, \href
  {https://ui.adsabs.harvard.edu/abs/2015RAA....15.2244Y} {15, 2244}

\bibitem[\protect\citeauthoryear{{Zhang} \& {Qian}}{{Zhang} \&
  {Qian}}{2020}]{2020MNRAS.497.3493Z}
{Zhang} X.-D.,  {Qian} S.-B.,  2020, \mn@doi [\mnras] {10.1093/mnras/staa2166},
  \href {https://ui.adsabs.harvard.edu/abs/2020MNRAS.497.3493Z} {497, 3493}

\makeatother
\end{thebibliography}


\clearpage
\setcounter{table}{0}
\begin{table*}
\caption{TESS minimum times of J069.}
\changefontsizes{8.5}
\centering
\begin{center}
\footnotesize
\begin{tabular}{c c c c c c c c c c c c}
 \hline
 \hline
Min.($BJD_{TDB}$) & Error & Epoch & O-C & Min.($BJD_{TDB}$) & Error & Epoch & O-C & Min.($BJD_{TDB}$) & Error & Epoch & O-C\\
\hline
2459910.27562	&	0.00186	&	6602	&	0.0144	&	2459919.08552	&	0.00036	&	6634.5	&	0.0146	&	2459927.90091	&	0.00032	&	6667	&	0.0163	\\
2459910.41298	&	0.00078	&	6602.5	&	0.0202	&	2459919.22223	&	0.00027	&	6635	&	0.0157	&	2459928.03478	&	0.00029	&	6667.5	&	0.0146	\\
2459910.54236	&	0.00062	&	6603	&	0.0140	&	2459919.35714	&	0.00026	&	6635.5	&	0.0151	&	2459928.17194	&	0.00028	&	6668	&	0.0162	\\
2459910.67939	&	0.00077	&	6603.5	&	0.0154	&	2459919.49317	&	0.00028	&	6636	&	0.0155	&	2459928.30612	&	0.00033	&	6668.5	&	0.0147	\\
2459910.81561	&	0.00053	&	6604	&	0.0160	&	2459919.62803	&	0.00033	&	6636.5	&	0.0148	&	2459928.44255	&	0.00039	&	6669	&	0.0156	\\
2459910.94952	&	0.00049	&	6604.5	&	0.0144	&	2459919.76422	&	0.00026	&	6637	&	0.0154	&	2459928.57705	&	0.00028	&	6669.5	&	0.0145	\\
2459911.08621	&	0.00041	&	6605	&	0.0154	&	2459919.89879	&	0.00031	&	6637.5	&	0.0143	&	2459928.71386	&	0.00028	&	6670	&	0.0157	\\
2459911.22140	&	0.00040	&	6605.5	&	0.0150	&	2459920.03584	&	0.00029	&	6638	&	0.0158	&	2459928.84887	&	0.00031	&	6670.5	&	0.0151	\\
2459911.35756	&	0.00036	&	6606	&	0.0156	&	2459920.17042	&	0.00026	&	6638.5	&	0.0148	&	2459928.98501	&	0.00025	&	6671	&	0.0157	\\
2459911.49217	&	0.00036	&	6606.5	&	0.0146	&	2459920.30681	&	0.00037	&	6639	&	0.0156	&	2459929.11922	&	0.00034	&	6671.5	&	0.0143	\\
2459911.62880	&	0.00036	&	6607	&	0.0157	&	2459920.44173	&	0.00038	&	6639.5	&	0.0149	&	2459929.25606	&	0.00032	&	6672	&	0.0155	\\
2459911.76355	&	0.00029	&	6607.5	&	0.0148	&	2459920.57792	&	0.00033	&	6640	&	0.0155	&	2459929.39087	&	0.00035	&	6672.5	&	0.0147	\\
2459911.90043	&	0.00036	&	6608	&	0.0161	&	2459920.71294	&	0.00032	&	6640.5	&	0.0149	&	2459929.52808	&	0.00048	&	6673	&	0.0163	\\
2459912.03441	&	0.00029	&	6608.5	&	0.0145	&	2459920.84930	&	0.00032	&	6641	&	0.0157	&	2459929.66143	&	0.00039	&	6673.5	&	0.0141	\\
2459912.17173	&	0.00036	&	6609	&	0.0162	&	2459920.98343	&	0.00067	&	6641.5	&	0.0142	&	2459929.79943	&	0.00051	&	6674	&	0.0165	\\
2459912.30597	&	0.00036	&	6609.5	&	0.0148	&	2459921.12003	&	0.00030	&	6642	&	0.0152	&	2459929.93359	&	0.00080	&	6674.5	&	0.0151	\\
2459912.44262	&	0.00025	&	6610	&	0.0159	&	2459921.25557	&	0.00027	&	6642.5	&	0.0152	&	2459930.07013	&	0.00101	&	6675	&	0.0160	\\
2459912.57741	&	0.00030	&	6610.5	&	0.0151	&	2459921.39154	&	0.00033	&	6643	&	0.0155	&	2459930.61128	&	0.00039	&	6677	&	0.0148	\\
2459912.71443	&	0.00038	&	6611	&	0.0165	&	2459921.52672	&	0.00039	&	6643.5	&	0.0151	&	2459930.74660	&	0.00042	&	6677.5	&	0.0145	\\
2459912.84806	&	0.00027	&	6611.5	&	0.0146	&	2459921.66269	&	0.00031	&	6644	&	0.0155	&	2459930.88272	&	0.00046	&	6678	&	0.0150	\\
2459912.98517	&	0.00030	&	6612	&	0.0161	&	2459921.79768	&	0.00031	&	6644.5	&	0.0149	&	2459931.01830	&	0.00030	&	6678.5	&	0.0150	\\
2459913.11928	&	0.00032	&	6612.5	&	0.0146	&	2459921.93380	&	0.00032	&	6645	&	0.0154	&	2459931.15473	&	0.00028	&	6679	&	0.0158	\\
2459913.25571	&	0.00036	&	6613	&	0.0154	&	2459922.06875	&	0.00035	&	6645.5	&	0.0148	&	2459931.28864	&	0.00035	&	6679.5	&	0.0142	\\
2459913.39003	&	0.00042	&	6613.5	&	0.0141	&	2459922.20518	&	0.00029	&	6646	&	0.0156	&	2459931.42555	&	0.00027	&	6680	&	0.0155	\\
2459913.52758	&	0.00039	&	6614	&	0.0161	&	2459922.33985	&	0.00036	&	6646.5	&	0.0147	&	2459931.56074	&	0.00027	&	6680.5	&	0.0151	\\
2459913.66140	&	0.00031	&	6614.5	&	0.0143	&	2459922.47612	&	0.00045	&	6647	&	0.0153	&	2459931.69710	&	0.00033	&	6681	&	0.0158	\\
2459913.79898	&	0.00037	&	6615	&	0.0163	&	2459922.61144	&	0.00039	&	6647.5	&	0.0151	&	2459931.83131	&	0.00030	&	6681.5	&	0.0144	\\
2459913.93310	&	0.00035	&	6615.5	&	0.0148	&	2459922.74779	&	0.00044	&	6648	&	0.0158	&	2459931.96807	&	0.00020	&	6682	&	0.0156	\\
2459914.06941	&	0.00026	&	6616	&	0.0155	&	2459922.88236	&	0.00045	&	6648.5	&	0.0148	&	2459932.10292	&	0.00029	&	6682.5	&	0.0149	\\
2459914.20499	&	0.00029	&	6616.5	&	0.0155	&	2459923.01795	&	0.00046	&	6649	&	0.0148	&	2459932.23949	&	0.00032	&	6683	&	0.0158	\\
2459914.34081	&	0.00034	&	6617	&	0.0158	&	2459923.15421	&	0.00044	&	6649.5	&	0.0155	&	2459932.37404	&	0.00027	&	6683.5	&	0.0148	\\
2459914.47544	&	0.00027	&	6617.5	&	0.0148	&	2459923.29005	&	0.00035	&	6650	&	0.0157	&	2459932.51086	&	0.00027	&	6684	&	0.0160	\\
2459914.61189	&	0.00047	&	6618	&	0.0156	&	2459923.69570	&	0.00059	&	6651.5	&	0.0146	&	2459932.64566	&	0.00022	&	6684.5	&	0.0152	\\
2459914.74663	&	0.00031	&	6618.5	&	0.0148	&	2459923.84099	&	0.00088	&	6652	&	0.0143	&	2459932.78154	&	0.00026	&	6685	&	0.0155	\\
2459914.88352	&	0.00034	&	6619	&	0.0161	&	2459923.96662	&	0.00091	&	6652.5	&	0.0143	&	2459932.91648	&	0.00032	&	6685.5	&	0.0149	\\
2459915.01813	&	0.00037	&	6619.5	&	0.0151	&	2459924.10261	&	0.00058	&	6653	&	0.0147	&	2459933.05245	&	0.00025	&	6686	&	0.0152	\\
2459915.15451	&	0.00034	&	6620	&	0.0159	&	2459924.23869	&	0.00065	&	6653.5	&	0.0152	&	2459933.18805	&	0.00030	&	6686.5	&	0.0152	\\
2459915.28901	&	0.00029	&	6620.5	&	0.0148	&	2459924.37433	&	0.00059	&	6654	&	0.0152	&	2459933.32419	&	0.00031	&	6687	&	0.0158	\\
2459915.42586	&	0.00026	&	6621	&	0.0160	&	2459924.50982	&	0.00045	&	6654.5	&	0.0151	&	2459933.45854	&	0.00030	&	6687.5	&	0.0145	\\
2459915.56049	&	0.00035	&	6621.5	&	0.0151	&	2459924.64557	&	0.00035	&	6655	&	0.0153	&	2459933.59560	&	0.00025	&	6688	&	0.0160	\\
2459915.69649	&	0.00032	&	6622	&	0.0155	&	2459924.78066	&	0.00042	&	6655.5	&	0.0148	&	2459933.72996	&	0.00030	&	6688.5	&	0.0148	\\
2459915.83174	&	0.00046	&	6622.5	&	0.0151	&	2459924.91758	&	0.00035	&	6656	&	0.0161	&	2459933.86626	&	0.00031	&	6689	&	0.0155	\\
2459915.96896	&	0.00044	&	6623	&	0.0168	&	2459925.05197	&	0.00035	&	6656.5	&	0.0149	&	2459934.00121	&	0.00032	&	6689.5	&	0.0148	\\
2459916.10182	&	0.00033	&	6623.5	&	0.0140	&	2459925.18868	&	0.00032	&	6657	&	0.0160	&	2459934.13743	&	0.00028	&	6690	&	0.0154	\\
2459916.23974	&	0.00037	&	6624	&	0.0164	&	2459925.32298	&	0.00028	&	6657.5	&	0.0147	&	2459934.27310	&	0.00032	&	6690.5	&	0.0155	\\
2459916.37408	&	0.00059	&	6624.5	&	0.0151	&	2459925.45961	&	0.00028	&	6658	&	0.0157	&	2459934.40837	&	0.00033	&	6691	&	0.0152	\\
2459916.50975	&	0.00070	&	6625	&	0.0152	&	2459925.59473	&	0.00030	&	6658.5	&	0.0153	&	2459934.54375	&	0.00033	&	6691.5	&	0.0150	\\
2459916.64458	&	0.00032	&	6625.5	&	0.0144	&	2459925.73115	&	0.00030	&	6659	&	0.0161	&	2459934.68009	&	0.00026	&	6692	&	0.0157	\\
2459917.05147	&	0.00034	&	6627	&	0.0145	&	2459925.86570	&	0.00036	&	6659.5	&	0.0150	&	2459934.81505	&	0.00027	&	6692.5	&	0.0151	\\
2459917.18709	&	0.00029	&	6627.5	&	0.0145	&	2459926.00200	&	0.00027	&	6660	&	0.0157	&	2459934.95130	&	0.00032	&	6693	&	0.0157	\\
2459917.32380	&	0.00031	&	6628	&	0.0156	&	2459926.13712	&	0.00035	&	6660.5	&	0.0153	&	2459935.08643	&	0.00027	&	6693.5	&	0.0153	\\
2459917.45834	&	0.00023	&	6628.5	&	0.0146	&	2459926.27298	&	0.00037	&	6661	&	0.0155	&	2459935.22264	&	0.00040	&	6694	&	0.0159	\\
2459917.59498	&	0.00037	&	6629	&	0.0156	&	2459926.40788	&	0.00033	&	6661.5	&	0.0148	&	2459935.35716	&	0.00036	&	6694.5	&	0.0148	\\
2459917.73143	&	0.00037	&	6629.5	&	0.0165	&	2459926.54431	&	0.00025	&	6662	&	0.0157	&	2459935.49349	&	0.00031	&	6695	&	0.0156	\\
2459917.86580	&	0.00026	&	6630	&	0.0153	&	2459926.67900	&	0.00028	&	6662.5	&	0.0148	&	2459935.62838	&	0.00035	&	6695.5	&	0.0148	\\
2459918.00101	&	0.00025	&	6630.5	&	0.0149	&	2459926.81575	&	0.00030	&	6663	&	0.0159	&	2459935.76411	&	0.00032	&	6696	&	0.0150	\\
2459918.13705	&	0.00037	&	6631	&	0.0153	&	2459926.95029	&	0.00034	&	6663.5	&	0.0149	&	2459935.89991	&	0.00035	&	6696.5	&	0.0152	\\
2459918.27234	&	0.00028	&	6631.5	&	0.0150	&	2459927.08661	&	0.00025	&	6664	&	0.0156	&	2459936.03614	&	0.00032	&	6697	&	0.0158	\\
2459918.40849	&	0.00031	&	6632	&	0.0156	&	2459927.22123	&	0.00031	&	6664.5	&	0.0146	&	2459936.17156	&	0.00031	&	6697.5	&	0.0156	\\
2459918.54285	&	0.00030	&	6632.5	&	0.0143	&	2459927.35828	&	0.00033	&	6665	&	0.0161	&	2459936.30694	&	0.00032	&	6698	&	0.0154	\\
2459918.68004	&	0.00024	&	6633	&	0.0159	&	2459927.49238	&	0.00033	&	6665.5	&	0.0146	&	2459936.44195	&	0.00031	&	6698.5	&	0.0148	\\
2459918.81486	&	0.00025	&	6633.5	&	0.0152	&	2459927.62959	&	0.00025	&	6666	&	0.0162	&	2459936.57899	&	0.00047	&	6699	&	0.0163	\\
2459918.95084	&	0.00028	&	6634	&	0.0155	&	2459927.76383	&	0.00031	&	6666.5	&	0.0148	&		&		&		&		\\
\hline
\hline
\end{tabular}
\end{center}
\label{A1}
\end{table*}

\begin{table*}
\caption{TESS minimum times of N581.}
\changefontsizes{8.5}
\centering
\begin{center}
\footnotesize
\begin{tabular}{c c c c c c c c c c c c}
 \hline
 \hline
Min.($BJD_{TDB}$) & Error & Epoch & O-C & Min.($BJD_{TDB}$) & Error & Epoch & O-C & Min.($BJD_{TDB}$) & Error & Epoch & O-C\\
\hline
2459825.28496	&	0.00030	&	7324.5	&	-0.0168	&	2459834.56004	&	0.00022	&	7362	&	-0.0127	&	2459843.95486	&	0.00026	&	7400	&	-0.0125	\\
2459825.41222	&	0.00025	&	7325	&	-0.0132	&	2459834.67913	&	0.00026	&	7362.5	&	-0.0172	&	2459844.07336	&	0.00030	&	7400.5	&	-0.0176	\\
2459825.53139	&	0.00029	&	7325.5	&	-0.0176	&	2459834.80690	&	0.00025	&	7363	&	-0.0131	&	2459844.20211	&	0.00021	&	7401	&	-0.0125	\\
2459825.65938	&	0.00030	&	7326	&	-0.0132	&	2459834.92631	&	0.00022	&	7363.5	&	-0.0173	&	2459844.32019	&	0.00027	&	7401.5	&	-0.0180	\\
2459825.77902	&	0.00028	&	7326.5	&	-0.0172	&	2459835.05414	&	0.00026	&	7364	&	-0.0131	&	2459844.44904	&	0.00022	&	7402	&	-0.0128	\\
2459825.90651	&	0.00023	&	7327	&	-0.0133	&	2459835.17326	&	0.00032	&	7364.5	&	-0.0176	&	2459844.56787	&	0.00028	&	7402.5	&	-0.0176	\\
2459826.02662	&	0.00024	&	7327.5	&	-0.0168	&	2459835.30138	&	0.00020	&	7365	&	-0.0131	&	2459844.69690	&	0.00029	&	7403	&	-0.0121	\\
2459826.15378	&	0.00025	&	7328	&	-0.0133	&	2459835.42061	&	0.00028	&	7365.5	&	-0.0174	&	2459844.81504	&	0.00033	&	7403.5	&	-0.0176	\\
2459826.27362	&	0.00029	&	7328.5	&	-0.0170	&	2459835.54863	&	0.00027	&	7366	&	-0.0130	&	2459844.94384	&	0.00029	&	7404	&	-0.0124	\\
2459826.40115	&	0.00025	&	7329	&	-0.0131	&	2459835.66745	&	0.00034	&	7366.5	&	-0.0178	&	2459845.06178	&	0.00026	&	7404.5	&	-0.0181	\\
2459826.52064	&	0.00031	&	7329.5	&	-0.0173	&	2459835.79586	&	0.00028	&	7367	&	-0.0130	&	2459845.19117	&	0.00025	&	7405	&	-0.0123	\\
2459826.64828	&	0.00026	&	7330	&	-0.0132	&	2459835.91508	&	0.00033	&	7367.5	&	-0.0174	&	2459845.30955	&	0.00023	&	7405.5	&	-0.0176	\\
2459826.76841	&	0.00026	&	7330.5	&	-0.0167	&	2459836.04406	&	0.00028	&	7368	&	-0.0121	&	2459845.43818	&	0.00022	&	7406	&	-0.0125	\\
2459826.89571	&	0.00025	&	7331	&	-0.0130	&	2459836.16164	&	0.00029	&	7368.5	&	-0.0181	&	2459845.55658	&	0.00025	&	7406.5	&	-0.0178	\\
2459827.01497	&	0.00030	&	7331.5	&	-0.0174	&	2459836.29055	&	0.00030	&	7369	&	-0.0128	&	2459845.68536	&	0.00022	&	7407	&	-0.0126	\\
2459827.14318	&	0.00030	&	7332	&	-0.0128	&	2459836.40948	&	0.00034	&	7369.5	&	-0.0175	&	2459845.80369	&	0.00030	&	7407.5	&	-0.0179	\\
2459827.26235	&	0.00031	&	7332.5	&	-0.0172	&	2459836.53744	&	0.00027	&	7370	&	-0.0131	&	2459846.18006	&	0.00025	&	7409	&	-0.0123	\\
2459827.39010	&	0.00022	&	7333	&	-0.0131	&	2459836.65685	&	0.00025	&	7370.5	&	-0.0173	&	2459846.29825	&	0.00027	&	7409.5	&	-0.0178	\\
2459827.50992	&	0.00030	&	7333.5	&	-0.0169	&	2459836.78471	&	0.00031	&	7371	&	-0.0131	&	2459846.42703	&	0.00028	&	7410	&	-0.0126	\\
2459827.63750	&	0.00026	&	7334	&	-0.0129	&	2459836.90343	&	0.00038	&	7371.5	&	-0.0180	&	2459846.54542	&	0.00031	&	7410.5	&	-0.0178	\\
2459827.75684	&	0.00026	&	7334.5	&	-0.0172	&	2459837.03270	&	0.00029	&	7372	&	-0.0123	&	2459846.67404	&	0.00026	&	7411	&	-0.0128	\\
2459827.88466	&	0.00023	&	7335	&	-0.0130	&	2459837.15085	&	0.00032	&	7372.5	&	-0.0178	&	2459846.79265	&	0.00026	&	7411.5	&	-0.0178	\\
2459828.00403	&	0.00033	&	7335.5	&	-0.0172	&	2459837.27921	&	0.00026	&	7373	&	-0.0130	&	2459846.92152	&	0.00025	&	7412	&	-0.0126	\\
2459828.13221	&	0.00029	&	7336	&	-0.0127	&	2459837.39791	&	0.00028	&	7373.5	&	-0.0180	&	2459847.03949	&	0.00029	&	7412.5	&	-0.0182	\\
2459828.25106	&	0.00024	&	7336.5	&	-0.0174	&	2459837.52640	&	0.00026	&	7374	&	-0.0131	&	2459847.16889	&	0.00023	&	7413	&	-0.0124	\\
2459828.37902	&	0.00024	&	7337	&	-0.0131	&	2459837.64557	&	0.00027	&	7374.5	&	-0.0175	&	2459847.28766	&	0.00031	&	7413.5	&	-0.0173	\\
2459828.49812	&	0.00023	&	7337.5	&	-0.0176	&	2459837.77404	&	0.00026	&	7375	&	-0.0127	&	2459847.41591	&	0.00026	&	7414	&	-0.0126	\\
2459828.62653	&	0.00023	&	7338	&	-0.0128	&	2459837.89236	&	0.00026	&	7375.5	&	-0.0180	&	2459847.53462	&	0.00029	&	7414.5	&	-0.0175	\\
2459828.74554	&	0.00032	&	7338.5	&	-0.0174	&	2459838.02087	&	0.00029	&	7376	&	-0.0131	&	2459847.66335	&	0.00020	&	7415	&	-0.0124	\\
2459828.87342	&	0.00028	&	7339	&	-0.0131	&	2459838.13992	&	0.00030	&	7376.5	&	-0.0176	&	2459847.78185	&	0.00029	&	7415.5	&	-0.0175	\\
2459828.99303	&	0.00028	&	7339.5	&	-0.0171	&	2459838.26838	&	0.00027	&	7377	&	-0.0128	&	2459847.91063	&	0.00021	&	7416	&	-0.0124	\\
2459829.12072	&	0.00024	&	7340	&	-0.0131	&	2459838.38629	&	0.00033	&	7377.5	&	-0.0185	&	2459848.02904	&	0.00030	&	7416.5	&	-0.0176	\\
2459829.23993	&	0.00029	&	7340.5	&	-0.0175	&	2459838.38629	&	0.00033	&	7377.5	&	-0.0185	&	2459848.15794	&	0.00020	&	7417	&	-0.0123	\\
2459829.36810	&	0.00028	&	7341	&	-0.0129	&	2459838.76273	&	0.00029	&	7379	&	-0.0129	&	2459848.27652	&	0.00026	&	7417.5	&	-0.0173	\\
2459829.48757	&	0.00026	&	7341.5	&	-0.0170	&	2459838.88144	&	0.00027	&	7379.5	&	-0.0178	&	2459848.40522	&	0.00024	&	7418	&	-0.0122	\\
2459829.61536	&	0.00017	&	7342	&	-0.0129	&	2459839.00997	&	0.00026	&	7380	&	-0.0129	&	2459848.52364	&	0.00028	&	7418.5	&	-0.0174	\\
2459829.73416	&	0.00029	&	7342.5	&	-0.0177	&	2459839.12865	&	0.00029	&	7380.5	&	-0.0178	&	2459848.65181	&	0.00026	&	7419	&	-0.0129	\\
2459829.86238	&	0.00025	&	7343	&	-0.0131	&	2459839.25718	&	0.00021	&	7381	&	-0.0129	&	2459848.77077	&	0.00031	&	7419.5	&	-0.0175	\\
2459829.98173	&	0.00029	&	7343.5	&	-0.0173	&	2459839.37628	&	0.00034	&	7381.5	&	-0.0174	&	2459848.89934	&	0.00027	&	7420	&	-0.0126	\\
2459830.10917	&	0.00024	&	7344	&	-0.0135	&	2459839.50454	&	0.00033	&	7382	&	-0.0128	&	2459849.01792	&	0.00022	&	7420.5	&	-0.0176	\\
2459830.22892	&	0.00032	&	7344.5	&	-0.0174	&	2459839.62298	&	0.00025	&	7382.5	&	-0.0179	&	2459849.14682	&	0.00025	&	7421	&	-0.0123	\\
2459830.35686	&	0.00024	&	7345	&	-0.0130	&	2459839.75168	&	0.00026	&	7383	&	-0.0128	&	2459849.26526	&	0.00030	&	7421.5	&	-0.0175	\\
2459830.47636	&	0.00029	&	7345.5	&	-0.0172	&	2459839.87023	&	0.00035	&	7383.5	&	-0.0179	&	2459849.39399	&	0.00026	&	7422	&	-0.0124	\\
2459830.60470	&	0.00023	&	7346	&	-0.0124	&	2459839.99847	&	0.00025	&	7384	&	-0.0133	&	2459849.51247	&	0.00033	&	7422.5	&	-0.0175	\\
2459830.72324	&	0.00029	&	7346.5	&	-0.0175	&	2459840.11766	&	0.00029	&	7384.5	&	-0.0177	&	2459849.64106	&	0.00028	&	7423	&	-0.0125	\\
2459830.85159	&	0.00028	&	7347	&	-0.0128	&	2459840.24615	&	0.00026	&	7385	&	-0.0128	&	2459849.75949	&	0.00028	&	7423.5	&	-0.0177	\\
2459830.97063	&	0.00027	&	7347.5	&	-0.0173	&	2459840.36495	&	0.00029	&	7385.5	&	-0.0176	&	2459849.88850	&	0.00025	&	7424	&	-0.0123	\\
2459831.09887	&	0.00021	&	7348	&	-0.0127	&	2459840.49312	&	0.00027	&	7386	&	-0.0131	&	2459850.00702	&	0.00027	&	7424.5	&	-0.0174	\\
2459831.21740	&	0.00028	&	7348.5	&	-0.0178	&	2459840.61225	&	0.00024	&	7386.5	&	-0.0176	&	2459850.13551	&	0.00031	&	7425	&	-0.0125	\\
2459831.34599	&	0.00021	&	7349	&	-0.0128	&	2459840.74059	&	0.00028	&	7387	&	-0.0128	&	2459850.25432	&	0.00025	&	7425.5	&	-0.0173	\\
2459831.46475	&	0.00025	&	7349.5	&	-0.0177	&	2459840.85932	&	0.00030	&	7387.5	&	-0.0177	&	2459850.38258	&	0.00030	&	7426	&	-0.0127	\\
2459831.59294	&	0.00027	&	7350	&	-0.0131	&	2459840.98839	&	0.00023	&	7388	&	-0.0123	&	2459850.50173	&	0.00031	&	7426.5	&	-0.0171	\\
2459831.71360	&	0.00055	&	7350.5	&	-0.0161	&	2459841.10628	&	0.00029	&	7388.5	&	-0.0180	&	2459850.62955	&	0.00025	&	7427	&	-0.0129	\\
2459831.71360	&	0.00055	&	7350.5	&	-0.0161	&	2459841.23525	&	0.00025	&	7389	&	-0.0126	&	2459850.74872	&	0.00026	&	7427.5	&	-0.0174	\\
2459831.96057	&	0.00038	&	7351.5	&	-0.0163	&	2459841.35383	&	0.00026	&	7389.5	&	-0.0177	&	2459850.87711	&	0.00028	&	7428	&	-0.0126	\\
2459832.08733	&	0.00026	&	7352	&	-0.0132	&	2459841.48242	&	0.00024	&	7390	&	-0.0127	&	2459850.99572	&	0.00025	&	7428.5	&	-0.0176	\\
2459832.20684	&	0.00023	&	7352.5	&	-0.0173	&	2459841.60086	&	0.00027	&	7390.5	&	-0.0179	&	2459851.12439	&	0.00023	&	7429	&	-0.0126	\\
2459832.33472	&	0.00026	&	7353	&	-0.0130	&	2459841.72962	&	0.00029	&	7391	&	-0.0127	&	2459851.24274	&	0.00029	&	7429.5	&	-0.0178	\\
2459832.45427	&	0.00028	&	7353.5	&	-0.0171	&	2459841.84820	&	0.00028	&	7391.5	&	-0.0177	&	2459851.37160	&	0.00031	&	7430	&	-0.0126	\\
2459832.58222	&	0.00022	&	7354	&	-0.0127	&	2459841.97685	&	0.00023	&	7392	&	-0.0127	&	2459851.49030	&	0.00028	&	7430.5	&	-0.0175	\\
2459832.70122	&	0.00023	&	7354.5	&	-0.0173	&	2459842.09542	&	0.00029	&	7392.5	&	-0.0178	&	2459851.61922	&	0.00029	&	7431	&	-0.0122	\\
2459832.82938	&	0.00026	&	7355	&	-0.0128	&	2459842.22420	&	0.00022	&	7393	&	-0.0126	&	2459851.73722	&	0.00030	&	7431.5	&	-0.0178	\\
2459832.94851	&	0.00032	&	7355.5	&	-0.0173	&	2459842.34282	&	0.00026	&	7393.5	&	-0.0176	&	2459851.86552	&	0.00027	&	7432	&	-0.0131	\\
2459833.07611	&	0.00026	&	7356	&	-0.0133	&	2459842.47132	&	0.00026	&	7394	&	-0.0127	&	2459851.98457	&	0.00026	&	7432.5	&	-0.0177	\\
2459833.19559	&	0.00026	&	7356.5	&	-0.0174	&	2459842.58978	&	0.00026	&	7394.5	&	-0.0178	&	2459852.11281	&	0.00021	&	7433	&	-0.0130	\\
2459833.32406	&	0.00024	&	7357	&	-0.0126	&	2459842.71872	&	0.00024	&	7395	&	-0.0125	&	2459852.23220	&	0.00028	&	7433.5	&	-0.0173	\\
2459833.44294	&	0.00031	&	7357.5	&	-0.0173	&	2459842.83712	&	0.00031	&	7395.5	&	-0.0177	&	2459852.36041	&	0.00023	&	7434	&	-0.0127	\\
2459833.57061	&	0.00025	&	7358	&	-0.0132	&	2459842.96546	&	0.00024	&	7396	&	-0.0130	&	2459852.47883	&	0.00023	&	7434.5	&	-0.0179	\\
2459833.69026	&	0.00023	&	7358.5	&	-0.0172	&	2459843.08423	&	0.00027	&	7396.5	&	-0.0178	&	2459852.60753	&	0.00019	&	7435	&	-0.0128	\\
2459833.81840	&	0.00022	&	7359	&	-0.0127	&	2459843.21275	&	0.00023	&	7397	&	-0.0129	&	2459852.72618	&	0.00025	&	7435.5	&	-0.0177	\\
2459833.93732	&	0.00027	&	7359.5	&	-0.0174	&	2459843.33154	&	0.00026	&	7397.5	&	-0.0178	&	2459852.85446	&	0.00028	&	7436	&	-0.0131	\\
2459834.06547	&	0.00023	&	7360	&	-0.0128	&	2459843.46022	&	0.00027	&	7398	&	-0.0127	&	2459852.97373	&	0.00021	&	7436.5	&	-0.0174	\\
2459834.18436	&	0.00033	&	7360.5	&	-0.0176	&	2459843.57855	&	0.00021	&	7398.5	&	-0.0180	&	2459853.10210	&	0.00021	&	7437	&	-0.0127	\\
2459834.31233	&	0.00024	&	7361	&	-0.0132	&	2459843.70759	&	0.00027	&	7399	&	-0.0126	&		&		&		&		\\
2459834.43169	&	0.00035	&	7361.5	&	-0.0175	&	2459843.82592	&	0.00029	&	7399.5	&	-0.0178	&		&		&		&		\\
\hline
\hline
\end{tabular}
\end{center}
\label{A2}
\end{table*}


\label{lastpage}
\end{document}